\documentclass[useAMS,usenatbib]{mn2e}
\usepackage{graphicx}
\usepackage{amssymb}
\usepackage{psfig}
\newcommand{\Lya}{Ly$\alpha$ }

\newcommand{\geqsim}{\,\raisebox{-0.6ex}{$\buildrel > \over \sim$}\,}
\newcommand{\leqsim}{\,\raisebox{-0.6ex}{$\buildrel < \over \sim$}\,}
\newcommand{\Kel}{ \, {\rm K}}
\newcommand{\Ang}{\, \mathring{ {\rm A}}}
\newcommand{\pc}{\, {\rm pc}}

\newcommand{\kmPs}{\,{\rm km}/{\rm s}}
\newcommand{\cm}{\, {\rm cm}}

\newcommand{\N}{\mathcal{N}}

\newcommand{\xesc}{x_{\rm e}}
\newcommand{\fesc}{f_{\rm e}}
\newcommand{\Tcr}{{T_4}}
\newcommand{\dd}{{\rm d}}

\newcommand{\EW}{{\rm EW}}
\newcommand{\xHI}{x_{\scriptscriptstyle {\rm HI}}}
\newcommand{\fT}{f^{\scriptscriptstyle {\rm T}}}

\begin{document}

\title{\Lya Radiative Transfer in a Multi-Phase Medium}   
\author[Hansen \& Oh]{Matthew Hansen \& S. Peng Oh\\ Dept. of Physics, University of California, Santa Barbara, CA 93106, USA.}
\maketitle

\begin{abstract}

Hydrogen Ly$\alpha$ is our primary emission-line window into high redshift galaxies. Surprisingly, despite an extensive literature, Ly$\alpha$ radiative transfer in the most realistic case of a dusty, multi-phase medium has not received detailed theoretical attention. We investigate Ly$\alpha$ resonant scattering through an ensemble of dusty, moving, optically thick gas clumps. 
We treat each clump as a scattering particle and use Monte Carlo simulations of surface scattering to quantify continuum and \Lya surface scattering angles, absorption probabilities, and frequency redistribution, as a function of the gas dust content. This atomistic approach speeds up the simulations by many orders of magnitude, making possible calculations which are otherwise intractable. Our fitting formulae can be readily adapted for fast radiative transfer in numerical simulations. With these surface scattering results, we develop an analytic framework for estimating escape fractions and line widths as a function of gas geometry, motion, and dust content. Our simple analytic model shows good agreement with full Monte-Carlo simulations. We show that the key geometric parameter is the average number of surface scatters for escape in the absence of absorption, $\N_0$, and we provide fitting formulae for several geometries of astrophysical interest.  We consider two interesting applications: (i) Equivalent widths. Ly$\alpha$ can preferentially escape from a dusty multi-phase ISM if most of the dust lies in cold neutral clouds, which Ly$\alpha$ photons cannot penetrate. This might explain the anomalously high EWs sometimes seen in high-redshift/submm sources. (ii) Multi-phase galactic outflows. We show the characteristic profile is asymmetric with a broad red tail, and relate the profile features to the outflow speed and gas geometry. Many future applications are envisaged. 
\end{abstract}

\section{Introduction}

The hydrogen Ly$\alpha$ line is our primary emission line window on the high-redshift universe. It is almost invariably crucial in securing redshift-identifications for the highest redshift galaxies 
(e.g., \citet{hu02a,hu02b,aji03,kod03,rho03,santosetal03}). Besides yielding redshifts, the shape of the line profile, equivalent width, and offset from other emission/absorption lines also encode information about the geometry, kinematics and underlying stellar population of the host galaxy. For instance, features in Ly$\alpha$ emission have been used to suggest strong galactic outflows \citep{kunthetal98}, as a signature of strong accretion shocks \citep{barkana_loeb}, and as evidence for an unusually strong ionizing continuum, perhaps due to Pop III stars \citep{malrhoads02}. Even after escaping the environs of the host galaxy, Ly$\alpha$ photons undergo processing in the surrounding intergalactic medium (IGM), and the presence or absence of observed Ly$\alpha$ emission can be used to place constraints on the epoch of reionization \citep{fan02,haiman02,santos03}. Because of the numerous factors which contribute to Ly$\alpha$ radiative transfer, the interpretation of such features is fraught with complexity. For instance, in a comprehensive set of $\sim 1000$ Lyman Break galaxies at $z\sim 3$, a plethora of Ly$\alpha$ strengths and line shapes were seen, ranging from pure damped absorption, to emission plus absorption, to pure strong emission \citep{shapley03}. Because of the tremendous potential returns for interpreting some rich data-sets, it is crucial to strive for a more detailed theoretical understanding of Ly$\alpha$ emission line features.

A very important factor in Ly$\alpha$ transmission is the presence of dust. Since the massive stars which produce metals evolve on a short timescale, and indeed supersolar metallicities \citep{pentericci} and CO emission \citep{bertoldi03} have been observed in the highest-redshift quasars at $z\sim 6$, dust is likely to be present in the ISM of even high-redshift galaxies. Because of their long scattering path-lengths, Ly$\alpha$ photons are extremely vulnerable to dust attenuation \citep{neufeld90,charlot_fall91}, and it was thought that this could account for low observed Ly$\alpha$ equivalent widths compared to that expected from optical Balmer emission lines \citep{meier,hartmann}, as well as early failures to detect high-redshift galaxies in blank sky surveys. However, further work has shown that dust content is not strongly correlated with Ly$\alpha$ equivalent width (where dust content can be inferred from metallicity or submillimeter emission). For instance, some dust-rich galaxies have significantly {\it higher} Ly$\alpha$ photon escape fractions than less dusty counterparts \citep{kunthetal98,kunthetal03}. Indeed, \citet{gia96} found a lack of correlation between the equivalent width of Ly$\alpha$ and the UV continuum slope $\beta$, which measures continuum extinction. They interpreted this as evidence for decoupling of the extinction of continuum and resonant line photons.
 
Such decoupling could take place if the interstellar medium is clumpy. \citet{neufeld91} and \citet{charlot_fall93} emphasized the importance of the geometry and multi-phase nature of the ISM in affecting the observed Ly$\alpha$ line. In particular, \citet{neufeld91} showed that in a clumpy, dusty ISM, the emergent Ly$\alpha$ emission could have a {\it higher} equivalent width than the unprocessed spectrum of the underlying stellar population. For instance, if the dust survives primarily in cold neutral clouds, Ly$\alpha$ photons scatter off the clouds and spend most of their time in the intercloud medium, whereas continuum photons propagate unhindered into the clouds and suffer greater extinction. Observationally, the ISM of our Galaxy is known to be clumpy down to small scales \citep{marscher,stutzki} , with a power-law cloud mass spectrum based on CO \citep{sanders} and 21cm \citep{dickey} emission data. From IRAS 100$\mu$m, CO and 21cm data, there is evidence for a multi-scale fractal structure for both the diffuse HI clouds \citep{bazell} as well as the molecular component \citep{elmegreen96}. The clumpiness of the ISM is well-established and it {\it must} be taken into account in radiative transfer calculations.  

Surprisingly, there has not been any detailed, quantitative, three-dimensional study of the effects of a dusty, clumpy, ISM on Ly$\alpha$ radiative transfer. The pioneering work of \citet{neufeld91} was a semi-analytic calculation for a plane-parallel slab: many issues, such as the detailed line profile and the effect of geometry, cannot be addressed with such an approach. Recently, \citet{richling} made a first attempt at a quantitative, three-dimensional calculation, but the slow convergence of the numerical technique employed restricted the study to line center optical depths of $\tau \le 100$, corresponding to neutral hydrogen column densities of $N \le 10^{16} \, {\rm cm^{-2}}$ for velocities $v \sim 100 \, {\rm km\, s^{-1}}$: orders of magnitude too low to be applicable to high-redshift galaxies. There have been many studies of the radiative transfer of UV continuum photons in a clumpy, dusty ISM, using a variety of techniques (e.g. \citet{witt_gordon,varosi_dwek,gordonetal01}), but none with extensions to resonance line photons.  Conversely,  while there have been Monte-Carlo radiative transfer studies of Ly$\alpha$ photons in both static media \citep{ahn01,ahn02} and expanding supershells \citep{ahn03}, all have only considered a uniform medium. This paper therefore represents a first attempt at numerically investigating Ly$\alpha$ radiative transfer incorporating both the effects of dust and gas clumping.

A key motivation is understanding recent puzzling observations of anomalous equivalent widths in high-redshift galaxies. For instance, high redshift $z=4.5,5.7$ sources observed by
\citet{rho03} in the Large Area Lyman Alpha survey (LALA) show anomolously large Ly$\alpha$ equivalent widths of EW$\ge 150 {\rm \AA}$
(rest-frame), many far in excess of any known nearby stellar
population. An AGN origin is
unlikely, as the observed upper limit on the X-ray to
Ly$\alpha$ ratio is about 4-24 times lower than the ratio for known
type II quasars \citep{malrhoads02}. The radiative transfer effects studied in this paper can produce an anomolously large Ly$\alpha$ equivalent width from a standard stellar population. Such an effect could also be at work in the mysterious Lyman-alpha emitters observed at $z\sim 3.1$ by \citet{steidel}, which have
enormous Ly$\alpha$ fluxes of $\sim 10^{-15} {\rm erg \, s^{-1}
  cm^{-2}}$ (a factor $\sim 20-40$ times large than typical line
emitters at the same redshift), but no observed continuum. Finally, our calculation could be of particular interest in interpreting the large ($\sim 1000$) sample of Ly-break galaxy spectra \citep{shapley03}, as well as understanding the spectra of galactic starbursts with winds.

The outline of this paper is as follows.  In \S\ref{section_scalings}, we derive the basic multiphase \Lya scaling relations.  We then consider radiative transfer off opaque gas surfaces in \S\ref{section_surfaces}, describing the Monte Carlo simulations and obtaining fitting formulae for the absorption probability, angular and frequency redistribution functions for both continuum and resonant scattering. With these surface scattering formulae in hand, we then develop a framework for multi-phase radiative transfer in \S\ref{section_mp}, where we derive escape fractions and \Lya line widths, discuss the role of the gas geometry, analyze several geometries of astrophysical relevance, and discuss the effects of dilute gas in between the opaque clumps. The surface scattering formulae substantially reduce the computational cost of simulating \Lya transfer, making otherwise intractable calculations feasible. We also develop a simple analytic model with a single geometric parameter that shows good agreement with the full Monte-Carlo simulations. In \S\ref{section:applications}, we discuss some applications of our formalism. We show how preferential absorption of continuum photons can lead to strong enhancement of the \Lya equivalent width. We also consider the typical \Lya line profiles resulting from outflows/inflows of multiphase gas, and relate the profile characteristics to the outflow/inflow speed and the gas geometry.  

\section{Scaling Relations for \Lya Absorption}\label{section_scalings}

In this section we build some physical intuition, by making simple order-of-magnitude estimates  for the absorption of Ly$\alpha$ photons in both homogeneous and multi-phase media, and summarizing some of the most important results from \S \ref{section_surfaces} and \S\ref{section_mp}. We shall see that for conditions prevailing in most galaxies, Ly$\alpha$ photons cannot escape unless the medium is multi-phase.  

Before beginning, it is useful to define some terms. Let $\nu_{\rm dop}=(V^{\rm dop}/c)\nu_{0}$ be the line Doppler width, where $\nu_0$ is the \Lya line center frequency, and $V^{\rm dop}=(2 k_{B}T/m_{p})^{1/2}$ is the characteristic atomic velocity dispersion times $\sqrt{2}$. We evaluate the frequency shift from line center in Doppler units, $x \equiv (\nu-\nu_{o})/\nu_{\rm dop}$\footnote{For gas at $10^4\Kel$ and a central frequency of $1216\Ang$, the frequency and wavelength conversions are: 1 Doppler width=12.85$\kmPs$=0.16$\Ang$.}. The Ly$\alpha$ scattering cross section is $\sigma(x)=\sigma_{0}\Phi(x)$, where $\Phi(x)$ is the Voigt function, which is characterized by a Gaussian Doppler core, and Lorentzian damping wings due to quantum broadening. For frequencies in the line wing, $|x| >3$, the Voigt function is dominated by the Lorentzian: $\Phi(x)\approx a/(\sqrt{\pi}x^2)$, where $a\equiv \nu_{\rm L}/2\nu_{\rm dop}=4.72\times10^{-4}\Tcr^{-1/2}$, $\nu_{\rm L}=4.03\times10^{-8}\nu_0$ is the width of the Lorenztian profile, and $T_{4}\equiv T/10^{4}$K. For ease of reference, we have listed the most common radiative transfer parameters used in this paper in Table \ref{table_parameters}. 
   
\begin{table}
\begin{center}
\begin{tabular}{| |p{3cm} | |p{0.25cm} | |p{3.75cm} |}
\hline
Parameter  & &  Value\\
\hline\hline
HI line center resonant scattering cross section & $\sigma_0$& $5.90\times10^{-14}{T_4}^{-1/2}\cm^2$\\\hline
Dust interaction cross section per hydrogen nucleus & $\sigma^d$&\\\hline
Dust absorption cross section per hydrogen nucleus & $\sigma^a$& $\epsilon_d \sigma^d$ \hbox{$\sigma^a_{-21}\equiv \sigma^a/10^{-21}\cm^2/H$}\\ \hline
Absorption parameter &$\beta$& $\sigma^a/\xHI\sigma_0$\\  \hline
Damping parameter &$a$& $\frac{\nu_{\rm L}}{2\nu_{\rm dop}}=4.72\times10^{-4}{T_4}^{-1/2}$\\ \hline
Frequency in doppler units &$x$& $\left(\nu-\nu_0 \right)/\nu_{\rm dop}$\\ \hline
Voigt function &$\Phi(x)$& $\quad\approx \frac{a}{\sqrt{\pi}x^2}\quad |x|\geqsim 3$\\ \hline
Doppler speed &$V^{\rm dop}$& $\sqrt{\frac{2k_b T}{m_H}}=12.85{T_4}^{1/2}\kmPs$\\ \hline
Absorption albedo &$\epsilon$&   $\frac{\sigma_{a}}{\sigma_{a} +\sigma_{s}}$ \\ \hline
Scattering asymmetry parameter &$g$& $\langle \cos\theta_{\rm scat}\rangle$ \\ \hline
\hline
\end{tabular}
\end{center}
\caption{{\bf Common Radiative Transfer Parameters} Note:  $T_4\equiv T/10^4\Kel$, $\xHI$ is the hydrogen neutral fraction, $\nu_0=2.48\times10^{15}{\rm s}^{-1}$ is the \Lya line center frequency, $\nu_{\rm dop}=(V^{\rm dop}/c)\nu_0$ is the doppler frequency, $\nu_{\rm L}=4.03\times 10^{-8} \nu_{0}$ is the width of the quantum broadening Lorentzian profile, and $\theta_{\rm scat}$ is the angle between the incident and outgoing photon directions.} 
\label{table_parameters}
\end{table}

\subsection{Homogeneous Slab}
\label{section:homogeneous_slab}

We begin by reviewing the physics of radiative transfer of Ly$\alpha$ photons through an optically thick slab, a problem that was first correctly solved by \citet{adams72}, and subsequently verified and explored in much greater detail \citep{harrington73,hummer80,bonilha79,Frisch80,neufeld90,ahn02}. We use these classical results to test our Monte-Carlo code in Appendix \ref{section_tests}. 

\subsubsection{Homogenous Slab:  Dust-Free}
Consider a \Lya photon escaping from a dust-free slab of pure HI with line center optical depth $\tau_0$. When the photon is the Doppler core, its mean free path is very short, and it barely diffuses spatially. It is always scattered by atoms with the same velocity along its direction of motion as the atom that emitted it. On rare occasions, it will encounter a fast moving atom in the tail of the Maxwellian velocity distribution, with large velocities perpendicular to the photon's direction. When this photon is re-emitted, it will be far from line-center, where the slab is optically thin. For a line-center optical depth of $\tau_{0}=10^{3}$, a frequency shift of $x \approx 2.6$ is sufficient to render the slab optically thin, $\tau \approx \tau_{0} e^{-x^{2}} \approx 1$, and the photon can escape. So escape from the medium is dominated by rare scattering events. 

However, if the medium is sufficiently optically thick, $\tau_{0} a > 10^{3}$, the non-negligible optical depth due to the damping wings still prevents escape. In this case, the photon will suffer repeated scatterings in the Lorentzian wings of typical atoms, and diffuse slowly in space and frequency, executing a random walk. Each scatter induces an r.m.s. Doppler shift of order $x\sim 1$, has a mean Doppler shift per scatter of $-1/|x|$ (with a bias to return to line center, due to the large probability for photons to scatter in the core
; \citet{Osterbrock62}), and transverses an optical depth $\tau_{0} \Phi(x) \Delta \tau \sim 1$, or a mean free path which is $\Delta \tau \sim 1/\Phi(x)$ line-center optical-depths. Hence, a photon at frequency $|x| \gg 1$ returns toward line center after $\N_{e} \sim x^{2}$ scatterings, having travelled an r.m.s. optical depth $\tau_{\rm rms} \sim \sqrt{\N_{e}} \Delta \tau \sim |x|/\Phi(x)$. If on its single longest excursion, the photon diffuses an r.m.s. distance of order the system size, $\tau_{\rm rms} \sim |x|/\Phi(x) \sim \tau_{0}$, then the photon can escape. Since $\Phi(x) \sim a/x^{2}$, this implies a critical escape frequency:
\begin{equation}
x_{e} = (a \tau_{0})^{1/3} \approx 30 \ T_{4}^{-1/3} N_{21}^{1/3},
\label{eqn:xescape}
\end{equation}
or almost $\sim 400 N_{21}^{1/3} {T_4}^{1/6}\ {\rm km \, s^{-1}}$ away from line center, where  $N_{21}\equiv N_{HI}/(10^{21} \, {\rm cm^{-2}})$. This displacement of photons away from line center can be seen in our Monte-Carlo simulations in Fig. \ref{slab_spectra}. 

\subsubsection{Homogeneous Slab:  Dusty}
Now let the gas contain dust, with an total (scattering + absorption) interaction cross section per hydrogen atom of $\sigma^d$, and an absorption probability per dust interaction $\epsilon_d$. The average absorption probability per interaction with either dust or hydrogen is:
\begin{eqnarray}\label{eqn:absorption_probability}
\epsilon&=&\frac{\sigma_{\rm absorb}}{\sigma_{\rm total}}=\frac{\epsilon_d\sigma^d}{\xHI \Phi(x)\sigma_0+\sigma^d}\\ \nonumber
&\approx& \frac{\beta}{\Phi(x)} \approx 1.59 \times 10^{-3} \ \frac{T_{4} \ \sigma_{-21}^{a}}{\xHI} \left[ \frac{x}{5}\right]^{2}\ ,
\end{eqnarray}
where $\xHI$ is the hydrogen neutral (HI) fraction (which must be introduced because $\sigma^d$ is the cross-section per hydrogen {\sl nuclei}).  In the third step we used the fact that, except very far from line center, HI scattering dominates, $\xHI \Phi(x)\sigma_0 \gg \sigma^d$, and defined the absorption parameter
\begin{equation}
\beta\equiv \epsilon_d\sigma^d/\xHI\sigma_0=1.69\times10^{-8} \ [T_4]^{1/2} [\xHI]^{-1}\sigma^{a}_{-21} \ ,
\end{equation}
where $\sigma^a\equiv \epsilon_d\sigma^d$. For the diffuse HI phase of the Milky Way, $ \sigma_{-21}^{a} \equiv \sigma^a/10^{-21}\cm^2/H\approx 1$, $\epsilon_d\approx 0.5$, and $\xHI\approx 1$ \citep{Draine84, Whittet03, Draine03}, and so $\beta \approx 10^{-8}$.  The fourth step in eqn. (\ref{eqn:absorption_probability}) uses the wing photon approximation $\Phi(x)\approx a/\sqrt{\pi} x^2$.

Under what conditions can the Ly$\alpha$ photon escape from such a dusty medium? While this has been the subject of detailed analytic and numerical work (e.g., \citet{hummer80, Frisch80, neufeld90}), we can understand the basic scaling laws quite easily. The probability that a photon will be absorbed at a given frequency $x$ is simply the number of scatterings at that frequency times the probability of absorption per scattering:
\begin{equation}
P_{\rm abs}^{\rm slab} \sim \N (x) \epsilon(x) \sim x^{2} \beta \frac{x^{2}}{a}. 
\label{eqn:Pabsorb}
\end{equation}
Thus, $P_{\rm abs}^{\rm slab} (x) \sim 1$ for
\begin{equation}
|x| > x_{\rm abs} \sim \left( \frac{a}{\beta} \right)^{1/4} \sim 12.9 \ \left[\frac{\xHI}{T_4 \ \sigma^a_{-21}}\right]^{1/4} \ . 
\label{eqn:xabsorb}
\end{equation}

This implies that a photon will be absorbed before escape if it has to diffuse far into the line wings in order to escape from the slab, or if 
\begin{eqnarray*}
x_{\rm e} > x_{\rm abs},
\end{eqnarray*}
with $x_{\rm e}$ and $x_{\rm abs}$ given by eqns. (\ref{eqn:xescape}) and (\ref{eqn:xabsorb}) respectively\footnote{Note that at $x_{\rm abs}$, the absorption probability per interaction is still small, $\epsilon \ll 1$, and scatterings still strongly predominate. The scattering and absorption cross-sections are comparable only at a much larger frequency, $x(\epsilon \sim 0.5) \sim (a/\beta)^{1/2} \sim x_{\rm absorb}^{2} \gg x_{\rm absorb}$, by which time all photons have been absorbed.}.  Hence, if the line center optical depth exceeds a critical value, 
\begin{equation}
\tau_{0} > \tau_{c} \approx \left( \frac{1}{a \beta^{3}} \right)^{1/4} \approx 4.6 \times 10^{6} \ T_{4}^{-1/4} [\xHI/\sigma_{-21}^{a}]^{3/4},
\end{equation} 
photons cannot escape from the medium. This simple criterion is borne out by more detailed calculations (e.g., see Fig. 5 of \citet{ahn00}, and references therein). In terms of the HI column density $N_{21}$, Ly$\alpha$ photons cannot escape from a homogeneous dusty slab once
\begin{equation}
N_{21} > 0.08 \ T_{4}^{1/2} [\xHI/\sigma_{-21}^{a}]^{3/4}.
\end{equation} 
Since typical HI column densities in the Milky Way and other galaxies is $N_{21} \sim 1$, Ly$\alpha$ photons could not escape if most of the HI is in a homogeneous dusty slab. In the next section, we see that if the gas is instead inhomogeneous/multi-phase, Ly$\alpha$ photons can escape much more easily. 

\subsection{Multiphase Gas}\label{section_mp_scalings}
We shall now estimate the absorption criteria for a multiphase dusty HI distribution. It is worth first noting that a medium is always more transparent when it is clumpy, for fairly generic and model independent reasons. The effective optical depth in an inhomogeneous medium is $\tau_{\rm clumpy}= -{\rm ln} \left( \langle {\rm exp}\left[-\tau \right] \rangle\right)$, where the average is over all lines of sight. However, for a uniform medium, $\tau={\rm constant}$ along all lines of sight, so that $\tau_{\rm uniform} = \langle \tau \rangle$. From the standard triangle inequality, 
\begin{equation}
\langle {\rm exp}\left[-\tau \right] \rangle \ \ge \ {\rm exp} \left[-\langle \tau \rangle \right] 
\end{equation}
and applying the negative logarithm to both sides, we see that $\tau_{\rm clumpy} \le \tau_{\rm uniform}$. Thus, for instance, flux transmission in quasar absorption spectra is increased for an inhomogeneous intergalactic medium (IGM), where transmission is dominated by underdense voids (e.g., \citet{fan02}). 

This effect is strongly exacerbated if most of the absorbing material lies in dense clumps which are optically thick to scattering. In this case, most of the photons scatter off the cloud surfaces without penetrating the clouds, which effectively shields the absorbing material. This situation naturally arises in a multi-phase ISM, when most of the dust lies in dense molecular/atomic clouds. For now, let us assume that the inter-cloud medium is highly ionized and relatively dust-free, so that all of the dust and HI lies in dense clouds. 

In \S \ref{section_SingleClouds}, we show that we can calculate analytically the escape probability of Ly$\alpha$ photons in a multi-phase medium quite accurately, given just two parameters: $\N_{0}$ and $\epsilon_c$. We define $\N_0$ as the mean number of cloud surfaces a photon would encounter before escape in the absence of absorption. It {\it only} depends on the geometry of the multi-phase medium (the trajectory of photons is independent of frequency, provided clouds are very optically thick). For most of the cases we will consider, $\N_0 \sim 1-30$, with typical values $\N_0 \sim 5$.  The cloud albedo $\epsilon_{c}$ is the probability of absorption upon hitting a cloud surface. In \S \ref{section_SingleClouds}, we show that: 
\begin{equation}
\epsilon_c\sim 2\sqrt{\epsilon}
\label{eqn:cloud_albedo}
\end{equation}
where $\epsilon$ is the absorption probability per interaction given by eqn. (\ref{eqn:absorption_probability}). Eqn. (\ref{eqn:cloud_albedo}) is easily understood in the case where $\epsilon$ is constant (e.g., for coherent scattering). The effective absorption optical depth of a medium with scattering is $\tau_{*} \approx \sqrt{\tau_{a} (\tau_{a}+\tau_{s})}$ (e.g., Rybicki \& Lightman, 1979, p. 38), where $\tau_a$ and $\tau_s$ are the absorption and scattering optical depths, respectively.  Hence, the albedo is $\epsilon_{c} = \tau_{*}/(\tau_{s}+\tau_{a}) \approx \sqrt{\tau_{a}/(\tau_a+\tau_{s})} = \sqrt{\epsilon}$. In \S \ref{section_SingleClouds}, we show this scaling still holds for Ly$\alpha$ photons, despite the fact that $\epsilon(x)$ changes as the photon random walks in frequency whilst scattering within the cloud.   We find that the typical frequency shift after scattering off an optically thick surface is $\Delta x\sim 1.5$, with most of the redistribution in a symmetric profile about the incident frequency $x_i$.  With $\epsilon(x)$ given by eqn. (\ref{eqn:absorption_probability}), the symmetric frequency distribution about $x_i$ implies $\left\langle \sqrt{\epsilon(x) } \right\rangle\approx \sqrt{\epsilon(\langle x\rangle) }\approx \sqrt{\epsilon( x_i )}$, since $\langle x \rangle = x_{i}$.  Therefore the coherent scattering absorption law describes the \Lya absorption when $\epsilon$ is evaluated at the incident frequency,  $\epsilon_c\sim 2\sqrt{\epsilon(x_i)}$.  We verify this explicitly in \S \ref{section_lya_absorption}.

Under the above approximations, the probability that a Ly$\alpha$ photon at frequency $x$ will be absorbed is:
\begin{equation}
P_{\rm abs}^{\rm multi-phase} \sim \N_0 \epsilon_{c} \sim 2 \N_0 \left( \frac{\beta}{a} \right)^{1/2} |x| 
\label{eqn:Pabsorb_MP}
\end{equation}
which should be compared against eqn. (\ref{eqn:Pabsorb}) for a homogenous slab. Notice the much weaker scaling with frequency: $P_{\rm abs}^{\rm multi-phase} \propto x$, instead of $P_{\rm abs}^{\rm slab} \propto x^{4}$. From eqn. (\ref{eqn:Pabsorb_MP}), $P_{\rm abs}^{\rm multi-phase} \sim 1$ for frequencies
\begin{eqnarray}
|x| > x_{\rm abs} &\approx&  \frac{1}{2 \N_0} \left( \frac{a}{\beta} \right)^{1/2} \\
&\approx& 16.7 \left[ \frac{\N_0}{5} \right]^{-1} \left[\frac{\xHI}{T_4 \ \sigma^a_{-21}}\right]^{1/2} \ .
\end{eqnarray}
Comparing against eqn. (\ref{eqn:xabsorb}), the cut-off absorption frequencies for the slab and multi-phase case are actually comparable, modulo the value of $\N_0$. However, there is an important difference: Ly$\alpha$ photons {\it have} to diffuse far into the wings in order to diffuse spatially out of an optically thick slab. Since escape requires $x_{\rm e} > x_{\rm absorb}$ for a very optically thick slab, the photons will inevitably be absorbed. By contrast, there is generally much less diffusion into the line wings when scattering off surfaces in a multi-phase medium. Photons typically only penetrate small optical depths, $\tau\sim 1-10$ in the cloud surfaces before escaping, and the number of scatterings is much less. Thus, the majority of photons need not necessarily stray far from line center. 

Clearly, the crucial parameter which determines if Ly$\alpha$ photons can escape from a multi-phase medium is $x_{\rm e}$, the characteristic escape frequency; this must be small, $x_{\rm e} < x_{\rm absorb}$, for photons to escape. Ly$\alpha$ photons in a multi-phase medium acquire Doppler frequency shifts in two ways: through the thermal motions of HI atoms, as before, and also through the bulk motions of the clouds/scattering surfaces. For this reason, it is useful to rewrite $x_{\rm abs}$ in units of velocity:
\begin{equation}
V^{\rm abs}_{2}= 2.1 \left[ \frac{\N_0}{5} \right]^{-1}  [\xHI/\sigma^{a}_{-21}]^{1/2}
\end{equation}
where $V_{2} \equiv V/100 \, {\rm km \, s^{-1}}$. In \S \ref{section_line_widths}, we show that atomic motions cause a net r.m.s. frequency shift of $\sim 1.5 \N_0 V^{\rm dop}$ after $\N_0$ surface scatterings, and consequently result in an escape frequency of:
\begin{equation}
V_{2}^{\rm e, atomic} \sim 0.5 V^{\rm dop}_2 \N_0  \sim 0.3 \ [T_{4}]^{1/2} \left[ \N_0/5 \right].  
\end{equation}
By contrast, we find that cloud motions (either random motions or bulk inflows/outflows) with characteristic velocity $V^{c}$ causes a net r.m.s. frequency shift of: 
\begin{equation}
V_{2}^{\rm e, cloud} \sim V^{c} \sqrt{\N_0} \sim 2.2 \  V^{c}_{2} \left[ \N_0/5\right]^{1/2}.
\end{equation}
Note that $V^{\rm e, atomic}\propto \N_0$, while $V^{\rm e, clouds}\propto \sqrt{\N_0}$, which we discuss in \S \ref{section_line_widths}.
When both the frequency redistribution and surface motion are combined, we find that the typical escape velocity is simply given by the sum (rather than the sum in quadrature), 
\begin{equation}\label{lya_absorb_V}
V^{\rm e}\sim V^{\rm e, atomic}+V^{\rm e, cloud}\ . 
\end{equation}

For absorption to be important, we require $V^{\rm e} > V^{\rm absorb}$, which constrains $\N_0$ to be
\begin{equation}
\N_0\geqsim  5 \ \left[0.38\left(T_4  \ \sigma^a_{-21}/\xHI\right)^{1/4}+\left( [V^c_2]^2 \sigma^a_{-21} /\xHI\right)^{1/3} \right]^{-1}\ .
\end{equation}
If $\N_0$ satisfies this inequality then the \Lya photons will be significantly absorbed.  This approximate constraint has the correct limits when either $V^{\rm e, atomic}=0$ or $V^{\rm e, cloud}=0$, but is off by a factor of $\sim 2$ when $V^{\rm e, atomic}\sim V^{\rm e, cloud}$.  The geometric parameter $\N_0$ thus plays a key role in determining if \Lya photons can escape in a multi-phase medium, and plays an analogous role to the column density $N_{\rm HI}$ in a homogeneous slab. In \S \ref{section_geometries}, we provide formulas for $\N_0$ for various different basic types of multi-phase geometries. 

\section{Surface Scattering \& Absorption}\label{section_surfaces}

Multi-phase radiative transfer typically involves photon propagation through an optically thin inter-cloud medium, and repeated scattering off optically thick clouds. In a full-blown Monte-Carlo simulation, the latter consumes by far the lion's share of computational time. This is extremely inefficient:  the same scattering/absorption problem off cloud surfaces is being solved over and over again for each photon. A better approach is to consider each cloud as a scattering/absorbing particle with its own radiative transfer properties (for other applications of this viewpoint, see \citet{neufeld91,hobson93,varosi_dwek}). We characterize these cloud scattering properties in this section.

For Ly$\alpha$ photons, clouds are extremely optically thick and have essentially the same radiative transfer properties as a semi-infinite slab. This eliminates detailed dependence on the geometry of the cloud:  all that matters is its dust content and the initial photon frequency. Surprisingly, the radiative transfer properties of a dusty semi-infinite slab to Ly$\alpha$ photon scattering has not been characterized in detail. We do so in this section. We derive formulas for the net absorption probability (the ``cloud albedo'') $\epsilon_c$, the exiting photon angular distribution $D(\theta)$, and the exiting photon frequency redistribution $R(x_i,x)$,
as a function of the initial photon frequency $x_i$ and the gas composition.  With these surface transfer formulae, radiative transfer through regions containing opaque gas clouds can be quickly estimated and/or simulated without performing any scattering calculations within the individual gas clouds.  This allows for both vast speed-ups of Monte-Carlo simulations (outlined in \S \ref{section_accelerated_mc}) and a tractable {\it analytic} multiphase radiative transfer analysis (\S \ref{section_mp}).
If a photon typically scatters $\N$ times before exiting a cloud (where $\N \sim 10^{5-7}$ for incident frequencies $x_i\sim 5$ and $\sigma^a_{-21}\sim 1$---see \S \ref{section_scatnum_Lya}),
then this allows a speed-up of order $\sim \N$, making tractable multi-phase calculations which would otherwise be prohibitively expensive.

Our approach is to find fitting formulas to Monte-Carlo simulations of an ensemble of incident photons.  Whenever possible, we base the fits on known analytic formulas for simple cases, extending the analytic formulas to encompass the more general cases that we simulate. We begin by describing the Monte-Carlo algorithm we use.  Surface radiative transfer of continuum photons, where scattering by dust is effectively coherent, is discussed next. We then consider the more complex case of \Lya surface scattering, where resonant frequency redistribution effects must be dealt with.  Lastly, we discuss two kinematic aspects of surface scattering:  the average scattering angle, and the frequency shift due to a bulk surface velocity.  

\subsection{Monte-Carlo Code Description}
The Monte Carlo algorithm we use is similar to the code used by \cite{ahn01, ahn02, ahn03} and \cite{zheng2}. These papers provide a fuller description of the algorithm than that which we give here.  An ensemble of photons is run through a medium with neutral HI and dust, 
and statistics are gathered.  Each photon is tracked until it either escapes the medium or is absorbed, at which point the photon's flight is terminated.  The optical depth $\tau$ between each interaction is drawn from the distribution $\exp(-\tau)$, i.e. $\tau= -\ln u$ where $u\in[0,1]$ is a random variable drawn from a uniform distribution (hereafter "univariate"), and the photon's position is updated.  The photon then interacts with either the HI or the dust, resulting in either HI resonant scattering, dust scattering, or dust absorption, all of which we describe next.  

We model the dust as particles which can either absorb or coherently scatter photons. Although dust scattering is not necessarily coherent, in practice ignoring frequency redistribution due to dust is an excellent approximation.  The trajectory of a continuum photon is unaffected by small deviations from coherent scattering, since the dust albedo $\epsilon_{d}$ only varies weakly with frequency.  However, the \Lya absorption probability is a strong function of frequency (see eqn. (\ref{eqn:absorption_probability}), so for resonant scattering the effects of frequency redistribution must be taken carefully into account.

A continuum photon only interacts with dust, with an absorption probability $\epsilon_d$ per interaction.  
We determine if the photon is absorbed during a given interaction by drawing a random variable $u\in[0,1]$ from a uniform distribution; if $u \le \epsilon_{d}$, the photon is deemed to be absorbed.  
If the continuum photon is not absorbed, then its direction is changed by the scattering angle $\theta_{\rm scat}$ off the incident direction, with a random azimuthal angle.  We use the Henyey-Greenstein scattering angle distribution\footnote{\citet{Draine03} shows that the Henyey-Greenstain distribution is inaccurate for wavelengths $\lambda\leq 4700\Ang$.  However, since the surface scattering problem we consider is essentially planar, the details of the dust scattering distribution should significantly affect the results.} (e.g., \citep{Witt77})
\begin{equation}\label{scattering_HG1}
P_{HG}(\theta;\,g_d)= \frac{1-{g_d}^2}{4\pi} \left(1+{g_d}^2-2g_d\,\cos\theta\right)^{-3/2}\ ,
\end{equation}
which is parameterized by the dust scattering asymmetry parameter $g_d\in[-1, 1]$, and where we use the normalization $1=\int_0^{\pi}{\rm d}\theta \ P(\theta)$. The scattering asymmetry parameter is defined as $g\equiv\langle \cos\theta_{\rm scat}\rangle$.  To approximate dust absorption and scattering in the Milky Way\citep{Draine84, witt_gordon, Whittet03, Draine03} at wavelengths near $1216\Ang$, we use $\epsilon_d=0.5$ and $g_d=0.5$, unless otherwise noted.  

If the photon is a \Lya photon, then the interaction can either be with dust or with neutral HI.  The probability of a dust interaction is $\sigma^d/(\sigma^d+\xHI\Phi(x)\sigma_0)$;  if a random univariate $u$ is less than this, then the interaction is identical to the ``continuum'' dust interaction described above.  Otherwise, the \Lya photon scatters resonantly off neutral hydrogen.  In all our simulations we take the hydrogen in the cold phase to be completely neutral, and so adopt $\xHI=1$ unless otherwise noted.  Although the velocity distribution of the hydrogen is Maxwellian, the velocity distribution of {\sl atoms that scatter photons} depends upon the frequency of the photon.  Let $\hat{n}$ be the direction of the photon before scattering.  In the two directions perpendicular to $\hat{n}$, the scattering atom's velocity distribution is Gaussian, 
\begin{equation}
f(w_{\perp})=\frac{1}{\sqrt{\pi}}e^{-{w_{\perp}}^2}
\end{equation}
where $w_{\perp}=v_{\perp}/V^{\rm dop}$ and $v_\perp$ is a velocity component in one of the two transverse directions to $\hat{n}$.   In the direction parallel to $\hat{n}$ the velocity distribution of {\sl scattering atoms} is a Gaussian weighted by a Lorentzian: the Gaussian is due to the thermal motion of the gas, while the Lorentzian is due to the increased probability for scattering in the wings from quantum mechanical broadening.  The velocity of the atom $v_{z}$ along the direction of the incident photon is determined by drawing a random variable from the distribution:
\begin{equation}
f(w_{z})=\frac{a}{\pi}\frac{{\rm exp}(-w_{z}^{2})}{(x_{i}-w_{z})^{2}+a^{2}} \frac{1}{\Phi(x_{i})}
\end{equation}
where $w_{z}=v_{z}/V^{\rm dop}$ (see \citet{zheng2} for a rapid algorithm for generating random numbers with this distribution) and $x_i$ is the photon's incident frequency.  In the rest frame of the atom, the frequency of the outgoing photon is the same as the incident frequency (strictly speaking it differs slightly due to the recoil effect \citep{field59}, but for our purposes this is negligible).  The new direction $\hat{n}'$ is given by a dipole distribution, with the symmetry axis defined by the incident direction $\hat{n}$:
\begin{equation}\label{dipole}
P(\theta)=\frac{3}{8}(1+\cos^2\theta)\ ,
\end{equation}
where $\theta$ is the polar angle off the direction $\hat{n}$.  Although resonant scattering can result in either isotropic or dipole scattering angle distributions, depending upon the intermediate excited quantum state \citep{stenflo80,ahn02}, the difference is immaterial for calculating spectra and escape fractions;  a more careful treatment would be required, for instance, to accurately simulate \Lya polarization.  Given the new photon direction $\hat{n}'$ and the scattering atom Doppler velocity $\vec{w}$, the new photon frequency $x'$ is given by
\begin{equation}
x'=x- \hat{n}\cdot \vec{w} + \hat{n}'\cdot \vec{w}\ .
\end{equation}

\subsubsection{Avoiding Core Scatters}\label{section_acceleration_scheme}
\Lya photons spend most of their scatters in the line core, where spatial diffusion is typically negligible.  Essentially, each time a photon enters the line core it scatters in place until it is scattered by a high speed atom which moves the frequency out of the core.  The frequency at the core-wing boundary, $x_{c}\approx 3$, is defined by:
\begin{equation}
\frac{a}{\sqrt{\pi} x_{c}^{2}} = \frac{1}{\sqrt{\pi}} e^{-x_{c}^{2}}.
\end{equation}
Hence, it typically takes $\exp({x_c}^2)\sim \exp(9)\sim 10^4$ scatters to scatter out of the core.  Note that $x_{c}$ depends only logarithmically on the gas temperature, through the damping parameter $a$. For a photon that starts out at frequency $x_i$ in the line wing, the photon typically returns to the core fairly quickly, after $\sim {x_i}^2 = 25 (x_{i}/5)^{2}$ scatters (see discussion in \S \ref{section:homogeneous_slab}).  Consequently, most of the simulation time is spent calculating core scatters.  By circumventing the core scatters, the simulation can be greatly sped up.  We have adopted a scheme to do so that is similar to that used by \cite{ahn02}, with the addition that we also consider absorption.  


One might think that absorption whilst scattering in the line core is negligible, due to the small physical path lengths traversed whilst scattering in the core. We confirm this quantitatively below. 

For a photon with an initial core frequency $x_i$ where $|x_i|<x_c$, let $\bar{n}_c$ be the average number of scatters for the photon to leave the core, $\tilde{x}_c \equiv \langle |x| \rangle$ be the average frequency (absolute magnitude) while in the core, and $\tilde{x}_w \equiv \langle |x_{w}| \rangle$ be the average frequency (absolute magnitude) of the first scatter that leaves the line core, $x_{w} > x_{c}$. We ran simulations for gas at $10^4\Kel$ and adopted $x_c=3$.  We find that for any initial frequency in the core, $\bar{n}_c\approx 2.9\times10^4$, $\bar{x}_c\approx 0.57$, and $\bar{x}_w\approx 3.3$ with an equal probability of leaving the core at $x=3.3$ and $x=-3.3$. Thus, the probability of absorption during core scatters can be approximated by (see eqn. (\ref{eqn:absorption_probability})):
\begin{eqnarray}
\epsilon_{\rm core}&\approx& \bar{n}_c \epsilon \approx \bar{n}_c \ \beta/\Phi(\bar{x}_c) \\
&\approx& 4.0 \times 10^{4} \beta= 6.8\times 10^{-4}  [\xHI]^{-1} \sigma^a_{-21}\nonumber \ .
\end{eqnarray}   
Since the typical photon scatters $\N\sim 10$ times before escaping the surface (\S \ref{section_ScatNumLya}) and it takes $\N^{\rm core}\sim 9(x_i/3)^2$ scatters to reach the core, a typical photon injected in the line wing will visit the core perhaps once.  The probability of a \Lya photon being absorbed in the core during the surface scattering is, therefore, $\sim 10^{-4}\sigma^a_{-21} (\epsilon_d/0.5) $, which is negligible.  


We therefore devised the following acceleration scheme\footnote{For completeness, this scheme still takes core absorption into account. This may be useful in other contexts when the core is revisited many times and core absorption could be non-negligible---e.g., for \Lya photons escaping from an optically thick slab.}: 1) If a photon starts off in the line core, we do the exact core scattering, and only employ the approximation scheme on subsequent visits to the core. This is computationally cheap, since an incident core photon does not penetrate deep into the surface, and typically leaves after a few scatters. 2)  If a photon enters the line core from the wing, the probability of absorption is $\epsilon_{\rm core}$. 3) If the photon is not absorbed, then it is given a wing frequency $x=\pm \tilde{x}_w$, with an equal probability for plus and minus. The spatial position of the photon is exactly the same as where it entered the line core, and the new angular direction is randomly drawn from an isotropic distribution.  In \S \ref{section_tests_exact} we compare this accelerated scheme to exact simulations of surface scattering. In practice, it gives accurate results, and gives a vast speed-up of the simulations, typically of order $\sim10^{5}$. 

\subsection{Surface Scattering of Continuum Photons}
In this section, we study the properties of coherent surface scattering of continuum photons.  It is very useful to understand the properties of coherent scattering surface transfer in order to have a baseline for comparison with \Lya surface transfer.  As such, in this section we do not use the Henyey-Greenstein scattering angle distribution, eqn. (\ref{scattering_HG1}), but instead use the same distribution that we use for \Lya scattering, which is the dipole distribution, eqn. (\ref{dipole})\footnote{Note that when we actually perform Monte-Carlo simulations of \Lya photons, we {\sl do} use the Henyey-Greenstein distribution when the \Lya photon scatters off dust.}.  We begin by calculating how thick a slab of gas must be before the surface scattering approximations apply.  We then describe fits for the cloud albedo $\epsilon_c$, the exiting photon scattering angle distribution $D(\theta)$, and the typical number of scatters $\N$.  

\subsubsection{The Surface Approximation}
\begin{figure}
\begin{center}
\vspace{0.5cm}
\scalebox{0.5}[0.5]{
\includegraphics[angle=0]{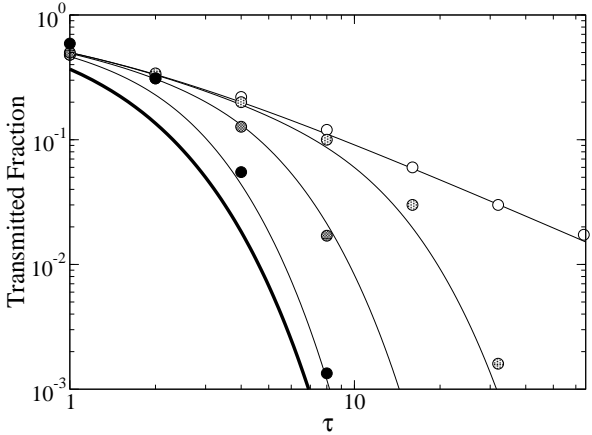} }
\caption{{\bf Transmission of Continuum Photons}  The fraction of incident photons that are transmitted through a finite slab, $\fT$, is shown as a function of the slab's total optical depth $\tau$, for various values of $\epsilon$. The circles are simulation results; from lightest to darkest, $\epsilon=0, \, 0.01, \, 0.1, \, 0.5$.  The thin lines are the fitting formula eqn. (\ref{transmit_fit}).  The thick line is the limiting case $\fT=\exp(-\tau)$, which corresponds to $\epsilon=1$.}
\label{figure_reflect}
\end{center}
\vspace{0cm}
\end{figure}

For a slab of material with a finite optical thickness $\tau$, the radiative transfer of photons incident on a surface will be approximately the same as for a semi-infinite slab if the fraction of photons that are transmitted through the slab, $\fT$, is small. In this limit, the surface is not translucent but acts as an absorbing mirror, and all photons are either reflected or absorbed. We define the penetration column density $N^{pt}$ such that when $N>N^{pt}$ the transmitted fraction is less than 10\%, $\fT<0.1$.  From a series of Monte-Carlo simulations, we find that a decent fitting formula for $f^t$ is  
\begin{equation}\label{transmit_fit}
\fT=\left[(1+\tau)\cosh\left(0.55\tau^{5/4}\epsilon^{1/2}\right)\right]^{-1}\ ,
\end{equation}
as shown in Figure \ref{figure_reflect}.  The transmission will be negligible ($\fT\leq 0.1$) when either $\tau\epsilon^{1/2}\geqsim 3$ or $\tau\geqsim 9$.   The corresponding penetration column density is
\begin{equation}
N^{pt}_{21}={\rm min}\left(
\frac{3}{\sigma^d_{-21}\sqrt{\epsilon}}, \ \frac{9}{\sigma^d_{-21}} \right) \ .    
\end{equation}

\subsubsection{Surface Absorption}
\begin{figure}
\begin{center}
\vspace{0.5cm}
\scalebox{0.5}[0.5]{
\includegraphics[angle=0]{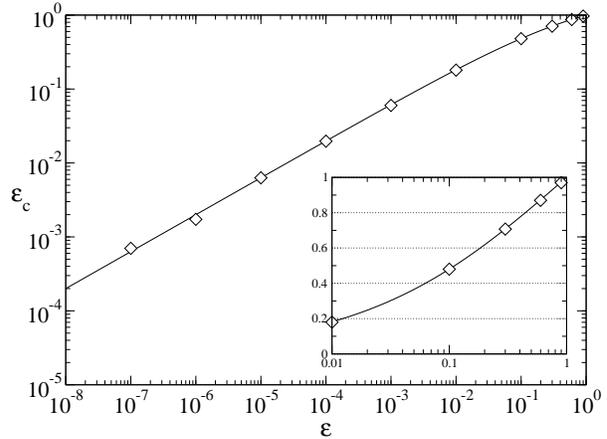}}
\caption{{\bf Continuum Photon Surface Absorption} 
Surface absorption probability $\epsilon_{c}$ for continuum photons, as a function of dust albedo $\epsilon$. The diamonds are simulations for coherent scattering with a dipole scattering angle distribution.  The line is the fit eqn. (\ref{albC_coherent}).}
\label{figure_albC}
\end{center}
\vspace{0cm}
\end{figure}

To derive a formula for the cloud albedo $\epsilon_c$, we ran simulations for an isotropic surface source and averaged the absorption over this ensemble.  For one dimensional radiative transfer, an exact formula for $\epsilon_c$ can be derived for photons incident on a semi-infinite line of material, 
\begin{equation}  \label{albC_coherent}
\epsilon_c=\frac{2\sqrt{\epsilon}}{1+\sqrt{\epsilon}}
\end{equation}
where each scatter is front-back symmetric ($g=0$)\footnote{For a plane-parallel slab, the Eddington approximation with the two-stream boundary condition gives $\fesc\propto \sqrt{\epsilon}/(1+\sqrt{\epsilon})$ (see, e.g., \citep{Rybicki79}, pg. 320), and from our simulations we find that the prefactor is $2$ for 1-D scattering.}.  
As shown by Figure \ref{figure_albC}, eqn. (\ref{albC_coherent}) provides a very good fit for the three dimensional, semi-infinite plane case.  When $\epsilon\ll 1$, we find $\epsilon_c\approx 2\sqrt{\epsilon}$, which is similar to the power law found by \cite{neufeld91}. 

\subsubsection{Escape Angles}\label{section_coh_scat}
\begin{figure}
\begin{center}
\vspace{0.5cm}
\scalebox{0.5}[0.5]{
\includegraphics[angle=0]{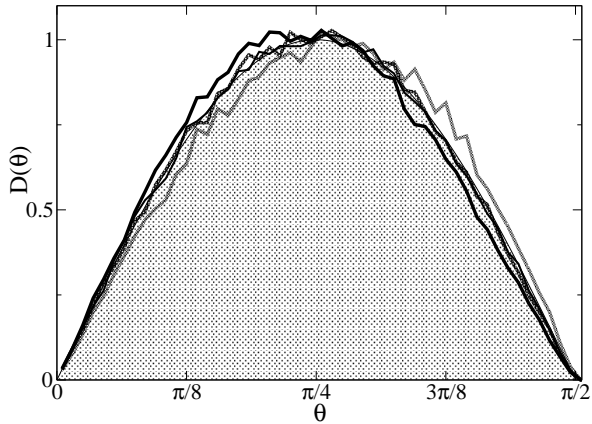} }
\caption{{\bf Exiting Angle Distribution}  The distribution of exiting angles relative to the surface normal is shown for several incident angles $\theta_{i}$ and absorption albedos $\epsilon$.  The shaded region denotes the distribution $D_{ss}(\theta)=\sin2\theta$.  The thick lines are dipole scattering: from darkest to lightest, the lines depict $(\epsilon,\theta_{i})=(10^{-4},0^o), \ (10^{-4},45^o), \ (0.1,45^o)$. The thin line is for isotropic scattering, $(\epsilon,\theta_{i})=(10^{-4},45^o)$.
}
\label{figure_surf_angles}
\end{center}
\vspace{0cm}
\end{figure}

To find a fit for the distribution of exiting angles for photons that escape, we ran simulations for various incident angles $\theta_{\rm i}$ relative to the surface normal.  We find that for nearly all $\theta_{\rm i}$, and for both isotropic and dipole single particle scattering, the distribution of exiting angles $\theta$ is well fit by
\begin{equation}\label{exit_angle}
D_{ss}(\theta)=\sin2\theta\ ,
\end{equation}
as shown in Figure \ref{figure_surf_angles}.  This distribution can be understood as the combination of two effects.  First, if photons scatter multiple times before escape, the photons effectively lose all ``memory'' of the incident angle.  This effect leads to a random exiting angle distribution, $D_{ss}(\theta)\propto \sin\theta$.  Second, photons that exit with angle $\theta$ will be attenuated if the optical depth traversed during the exiting leg, $\tau$, exceeds unity.   Let $\tau^\perp$ be the perpendicular optical depth at the point of last scatter for a photon that would exit with an angle $\theta$ in the absence of absorption.  The condition $\tau\leq 1$ implies a maximum perpendicular depth of such a photon is $\tau^\perp_{\rm max}=\cos\theta$ (e.g., only shallow surface layers contribute photons escaping nearly parallel to the surface).  Near the surface, the mean intensity is approximately constant for $\tau^\perp\leqsim 1$.  This implies that the number of photons available to escape at an angle $\theta$ scales as $\propto \tau^\perp_{\rm max}$, which implies $D_{ss}(\theta)\propto \cos\theta$.  Including both the effect of randomization and attenuation gives a distribution $D_{ss}(\theta)\propto \sin\theta \cos\theta$, which, when normalized over $\theta\in[0, \pi/2]$, gives eqn. (\ref{exit_angle}).  In the case of dipole scattering there are deviations from this fit for grazing incident angles, but overall this fit is generic for surface scattering when the single particle scattering distribution is front-back symmetric.  We also find that there is very little dependence on the absorption albedo $\epsilon$.  When the distribution $D_{ss}(\theta)$ holds, the distribution of azimuthal angles is uniform over the interval $[0, 2\pi)$.  Finally, we note that if the slab is viewed at an angle $\theta_{\rm obs}$ (from the outward surface normal), then the observed intensity is $\propto D_{ss}(\theta_{\rm obs})\cos\theta_{\rm obs}$.  The extra $\cos\theta_{\rm obs}$ factor is due to the dependence of the projected surface area on the viewing angle.  

\subsubsection{Number of Scatters for Escape}\label{section_scatnum_Lya}
\begin{figure}
\begin{center}
\vspace{0.5cm}
\scalebox{0.5}[0.5]{
\includegraphics[angle=0]{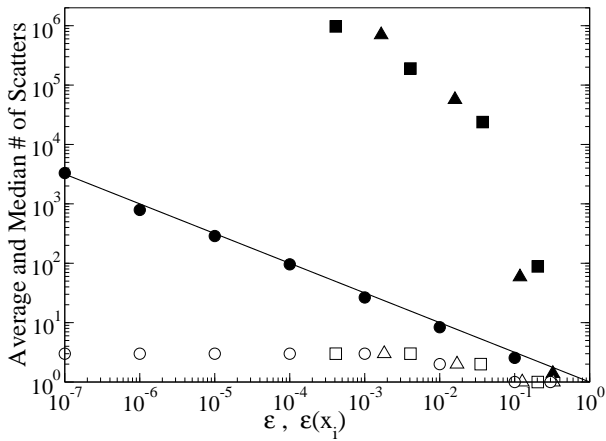} }
\caption{{\bf Number of Scatters: Continuum \& \Lya Photons}  The number of scatters for escape for coherently scattered photons (circles) and \Lya photons (squares and triangles) as a function of the absorption albedo.  For \Lya photons the absorption albedo is evaluated at the incident frequency;  two initial frequencies are shown, $x_i=10$ (squares) and $x_i=20$ (triangles).  For each case, both the average (filled symbols) and median (open symbols) number of scatters are shown.  The analytic estimate for the number of scatters for coherent scattering (solid line), given by eqn. (\ref{Nscat_coherent}), matches the data very well.}
\label{figure_scatnum}
\end{center}
\vspace{0cm}
\end{figure}

In \S \ref{section_analytic_coherent} we present a general derivation of the average number of scatters, $\N$, as a function of the escape fraction $\fesc$ and the absorption albedo $\epsilon$, given by eqn. (\ref{Nesc}).  Application of this formula to surface scattering, where the escape fraction is $\fesc=1-\epsilon_c(\epsilon)$, gives
\begin{equation}\label{Nscat_coherent}
\N= 1/\sqrt{\epsilon}\ ,
\end{equation}
which is shown by the solid line in Figure \ref{figure_scatnum}.  As the absorption albedo goes to zero, the average number of scatters diverges, although the median number of scatters does not seem to increase beyond $\sim 5$.  Clearly, the average is dominated by the rare photons that wander deep into the surface.

\subsection{Surface \Lya Transfer}\label{section_SingleClouds}
As in the case of coherent scattering considered above, we begin by calculating how thick a finite slab of gas must be before the surface scattering approximations apply.  We then describe fits for the net absorption $\epsilon_c$, scattering angle distribution $D(\theta)$, the typical number of scatters, and the Ly$\alpha$ frequency redistribution $R(x_i,x)$ as a function of the incident frequency $x_{i}$. 

\subsubsection{The \Lya Surface Approximation}\label{section_Npt_lya}
\begin{figure}
\begin{center}
\vspace{0.5cm}
\scalebox{0.5}[.5]{
\includegraphics[angle=0]{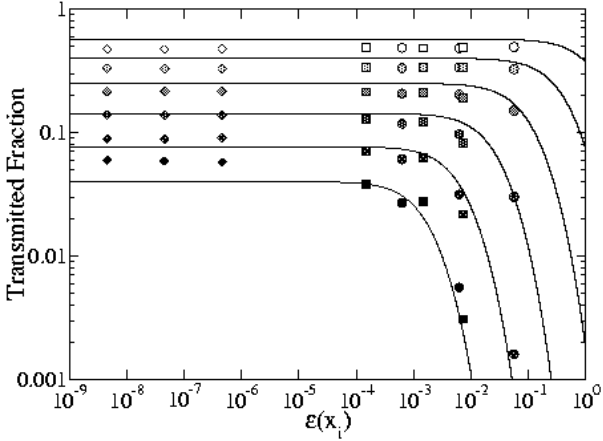} }
\caption{{\bf Transmission of \Lya Photons}  The fraction of escaping photons that are transmitted through the slab $\fT$ as a function of $\epsilon_i=\epsilon(x_i)$, for gas at $T=10^4\Kel$ with dust parameters $g_d=0.5$ and $\epsilon_d=0.5$.  Each shade is a different incident slab optical depth $\tau_i$:  from lightest to darkest $\tau_i=1, \, 2,\,  4,\,  8,\,  16,\,  32$.  Simulation data for three different incident frequencies $x_i$ are shown:  the diamonds, squares, and circles are $x_i=1, \, 5, \, 10$, respectively.  For each $x_i$ and $\tau_i$, simulations were run for three different dust absorption cross-sections:  reading from left to right for each $x_i$ and $\tau_i$, the dust values are $\sigma^a_{-21}=0.1, \, 1, \, 10$.  The lines are generated from the fitting formula eqn. (\ref{Xt_fit_lya}).}
\label{figure_reflect_lya}
\end{center}
\vspace{0cm}
\end{figure}

The criterion for the surface scattering approximation to apply for incident \Lya photons is similar to that defined for coherent scattering.  Consider \Lya photons with frequency $x_i$ incident on a finite slab with optical thickness $\tau_i=\tau_0\Phi(x_i)$, evaluated at the incident frequency.  When $x_i$ is in the Lorentzian wing and if the slab is pure HI,  \cite{neufeld90} analytically derived that the fraction transmitted is:
\begin{equation}
\fT_{\rm HI}=\frac{4}{3\tau_i}
\end{equation}
when $\tau_i\gg1$.  Our simulations confirm this result when there is very little dust, but find that $\fT$ can be substantially less than $\fT_{\rm HI}$ when a small amount of absorbing dust is present.  To derive a fitting formula for $\fT$ we ran simulations for various incident frequencies and dust absorption cross-sections.  To a large degree, $\fT$ depends only upon the incident single scattering albedo $\epsilon_i$ and the incident slab optical depth $\tau_i$;  almost all the frequency dependence is captured by these two parameters. This is extremely convenient: the transmitted fraction (and as we shall subsequently see, the reflected fraction) depends in a fairly simple way on the properties of the slab and the incident frequency. One might worry that due to frequency redistribution (which can be substantial; see \S \ref{section_surface_freqred}), the frequency dependence becomes extremely complicated, but that appears not to be the case. In particular, the typical incident photon never ``loses memory'' of its initial frequency. 
Most photons that escape do so after a handful of scatters, $\sim 10$.  When $x_i\geqsim 4$, the majority of photons do not wander into the core before escaping, and so most photons retain some memory of the incident frequency.
As shown in Figure \ref{figure_reflect_lya}, a reasonable fit for photons initially in the line wing is:
\begin{equation}\label{Xt_fit_lya}
\fT=\left[\left(1+\frac{3\tau_i}{4}\right)\cosh\left( \left(\tau_i\sqrt{\epsilon_i}\right)^{5/4}\right)\right]^{-1}\ .
\end{equation}
When $\epsilon_i\rightarrow 0$, the pure HI formula $\fT_{\rm HI}$, eqn. (\ref{transmit_fit}),  is recovered when $\tau_i\gg1$.  As can be seen in the figure, for photons initially in the Doppler core the transmitted fraction is slightly larger than this.  The transmission will be negligible if either $\tau_i\sqrt{\epsilon_i}\geqsim 3$ or $\tau_i\geqsim 12$.  This corresponds to a penetration column density
\begin{equation}\label{Npt_lya}
N^{pt}_{21}=
{\rm min} \left( 0.2 \ V_2 [\sigma^a_{-21}]^{-1/2}, \ 
0.05\ [V_2]^2 \right) 
\end{equation}
for any incident frequency $|x_i|\geq 3$, with $N^{pt}_{21}$ substantially smaller for photons in the Doppler core. 

\subsubsection{Surface \Lya Absorption}\label{section_lya_absorption}
\begin{figure}
\begin{center}
\vspace{0.5cm}
\scalebox{0.5}[0.5]{
\includegraphics[angle=0]{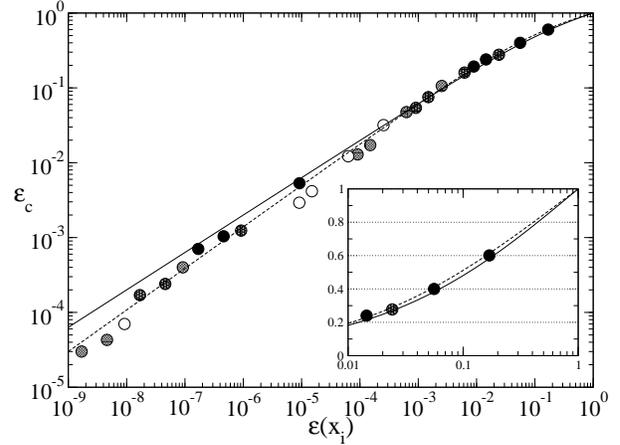}}
\caption{{\bf Ly$\alpha$ Surface Absorption (1)} The surface albedo $\epsilon_c$ is plotted against the single scattering albedo evaluated at the incident frequency, $\epsilon_i$, for gas at $T=10^4\Kel$ with dust parameters $g_d=0.5$ and $\epsilon_d=0.5$. The inset shows an enlargement of the $\epsilon=0.01-1$ region. Simulations were run for four different dust absorption cross-sections and seven different incident frequencies.  Each shade corresponds to a different $\sigma^a$:  from lightest to darkest $\sigma^a_{-21}=0.01,\ 0.1,\,  1,\,  10$.  For each value of $\sigma^a$, seven different incident frequencies $x_i$ were run: from right to left, $x_i=20, \, 10, \, 5, \, 4, \, 2, \, 1, \, 0$.  Note that for $x_i=0$, the value of $\epsilon_i$ for $\sigma^a_{-21}=0.01$ and $\sigma^a_{-21}=0.1$ is less than $10^{-9}$, and hence lies off the plot. The absorption probability $\epsilon_{c}$ depends only on $\epsilon_{i}$; at fixed $\epsilon_{i}$, there is no independent variation with dust content $\sigma_{-21}^{a}$ or incident frequency $x_{i}$. The solid line is the fit eqn. (\ref{albC_lya}), while the dashed line is the fit eqn. (\ref{albC_lya2}).}
\label{figure_albL}
\end{center}
\vspace{0cm}
\end{figure}

To derive a fit for the surface albedo $\epsilon_c$,  we ran again simulations for a wide range of incident frequencies and dust cross-sections.  As in the coherent scattering case, we average $\epsilon_c$ over an isotropic incident direction.  We find that $\epsilon_c$ mainly depends upon the single scattering albedo at the incident frequency, $\epsilon_i$.  As shown by the solid line in Figure \ref{figure_albL}, $\epsilon_c$ is well fit by
\begin{equation}\label{albC_lya}
\epsilon_c\approx \frac{2\sqrt{\epsilon_i}}{1+\sqrt{\epsilon_i}}\ ,
\end{equation}
which has the same form as for coherently scattered photons, eqn. (\ref{albC_coherent}).  A slightly better fit is shown by the dashed line in Figure \ref{figure_albL}, 
\begin{equation}\label{albC_lya2}
\epsilon_c\approx \frac{3 {\epsilon_i}^{5/9}}{1+2{\epsilon_i}^{1/2}}\ .
\end{equation}
Since eqn. (\ref{albC_lya2}) gives a slightly better fit only when $\epsilon_c$ is negligibly small, in practice we always use eqn. (\ref{albC_lya});  this will simplify the subsequent multiphase analysis somewhat, for only a small loss in accuracy.  Either of these formulas for $\epsilon_c$ applies even when $x_i$ is in the Doppler core (although eqn. (\ref{albC_lya2}) gives the better fit in this case).

\begin{figure}
\begin{center}
\vspace{0.5cm}
\scalebox{0.5}[0.5]{
\includegraphics[angle=0]{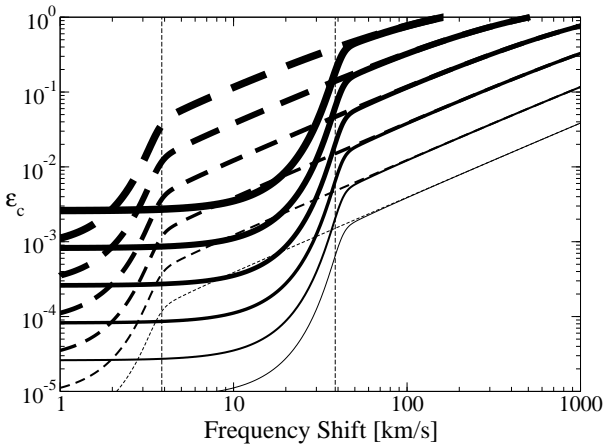} }
\caption{{\bf Ly$\alpha$ Surface Absorption (2)}  Simulation results for the net probability of surface absorption plotted as a function of the incident \Lya frequency shift off line center, in velocity units.  The solid lines are for a gas temperature $T=10^4\Kel$, while the dashed lines are for $T=10^2\Kel$.  We also show the effect of varying the dust content:  from the thickest line to the thinnest, $\sigma^{a}_{-21}=100, \, 10, \, 1, \, 0.1, \, 0.01, \, 0.001$. The cloud albedo $\epsilon_{c}$ is independent of gas temperature if the photon is in the line wing,  though the core-wing boundary is obviously temperature-dependent:  $V_c\approx 3V^{\rm dop}\propto T^{1/2}$ (the vertical dashed lines show the core-wing boundary at $T=100\Kel, \ 10^4\Kel$).
The lines flatten out once the incident frequency shift lies within the Doppler core, because the \Lya scattering cross section approaches its line center value $\sigma_0$.  This causes the single scattering albedo to asymptote as $V\rightarrow 0$.  Since $\epsilon_c\propto \sqrt{\epsilon}$ approximately holds even for incident photons in the Doppler core, the surface albedo will also asymptote as $V\rightarrow 0$.}
\label{figure_ec}
\end{center}
\vspace{0cm}
\end{figure}

As long as the Ly$\alpha$ photon is in the line wing, the absorption probability is independent of the gas temperature. To show this, in Figure \ref{figure_ec} we calculate $\epsilon_c$ using eqn. (\ref{albC_lya}), as a function of the incident velocity shift $\Delta V$ in physical units, rather than Doppler units $x$.  The figure shows the calculation for gas at temperatures $T=100\Kel$ and $T=10^4\Kel$ and for varying dust content $\sigma^a_{-21}$.  For photons in the line wing, the temperature dependence of $\epsilon$ drops out:
\begin{equation}
\epsilon^{\rm wing}_c\approx \frac{\sigma^a}{\xHI\Phi(x)\sigma_0}\approx \frac{\sqrt{\pi}x^2\sigma^a}{a\xHI\sigma_0}\propto\frac{x^2}{a\xHI\sigma_0}\ .
\end{equation}
Focusing just on the temperature, $a\propto 1/\sqrt{T}$, $\sigma_0\propto 1/\sqrt{T}$, and $x=V/V^{\rm dop}\propto 1/\sqrt{T}$.  Therefore the combination $x^2/a\sigma_0$ has no dependence on the gas temperature.  Physically, scattering in the Lorenztian wing is dominated by the quantum broadening of the cross-section, which does not depend upon the thermal motion of the scattering atoms.  Since we have shown that $\epsilon_c$ mainly depends upon $\epsilon_i$, it follows that $\epsilon_c$ is also temperature independent.  Thus, the surface absorption probability does not depend upon the gas temperature
 
\subsubsection{Escape Angles}
As argued in \S \ref{section_coh_scat} the scattering distribution $D_{ss}(\theta)$, eqn. (\ref{exit_angle}) is fairly generic when the single particle scattering is front-back symmetric ($g=0$).  Since \Lya scattering---which is either isotropic or dipole---is always front-back symmetric, the \Lya surface scattering angle distribution is also given by $D_{ss}(\theta)$, independent of the incident angle $\theta_{i}$. 

\subsubsection{Number of Scatters for Escape}\label{section_ScatNumLya}
The average number of scatters for \Lya photons to escape is more complicated than for continuum photons mainly because \Lya can be trapped in the Doppler core.  As discussed in \S \ref{section_acceleration_scheme}, \Lya photons in the Doppler core must scatter $\sim 3\times 10^4$ times before a rare scattering event brings the frequency into the line wing.  In Figure \ref{figure_scatnum} we show exact Monte-Carlo simulations of \Lya photons, where the core scatters are directly calculated, i.e. the Monte-Carlo acceleration scheme described in \S \ref{section_acceleration_scheme} is {\sl not} used.   The huge increase in scatterings over the continuum scattering result is obviously due to scattering in the Doppler core.  The average number of scatterings in the line wing is (analogous to the continuum scattering result)
\begin{equation}
\N^{\rm w}\sim 1/\sqrt{\epsilon(x_i)} \sim 25 \ [\xHI/\sigma^a_{-21}]^{1/2}\left[x_i/5\right]^{-1}\ ,
\end{equation}
while the average number of scatterings required to reach the Doppler core is $\sim 25 [x_i/5]^2$ (see the discussion in \S \ref{section_mp_scalings}).  Thus, although the typical photon does not reach the core, the probability of reaching the core  is large enough that the the {\sl average} $\N$ is dominated by core scattering.

\subsubsection{Surface \Lya Frequency Redistribution}\label{section_surface_freqred}
In this section, we find a formula for the reflected frequency distribution $R(x_i,x)$ as a function of the incident frequency $x_i$, for the pure dust-free HI case.  When dust is added, we find that the distribution $R_{\rm dust}(x)$ adheres closely to $R(x_i, x)$ as long as $\sigma^a_{-21}\xHI \leqsim 10$. Dust will have little effect on frequency redistribution, except for extremely dusty or highly ionized clouds.

For photons incident at frequency $x_i$ on an optically thick slab, $a\tau_0\geqsim 10^3$, an analytic solution for the transmitted and reflected emission profile has been derived by \cite{neufeld90}, extending the earlier work of \cite{harrington73} 
who obtained these results for the case $x_i=0$.  By taking the $\tau_0\rightarrow \infty$ limit of eqn. (2.33) in \cite{neufeld90}, we derive the analytic result for the reflected spectrum (normalized to unity):
\begin{equation}\label{R_neuf}
R_{\rm anlyt}(x_i,x)=\sqrt{\frac{3}{2\pi^2}} \frac{x^2{x_i}^2}{\left(x^3-{x_i}^3\right)^2/6+{x_i}^4}\ .
\end{equation}   
\begin{figure}
\begin{center}
\vspace{0.5cm}
\scalebox{0.4}[0.4]{
\includegraphics[angle=0]{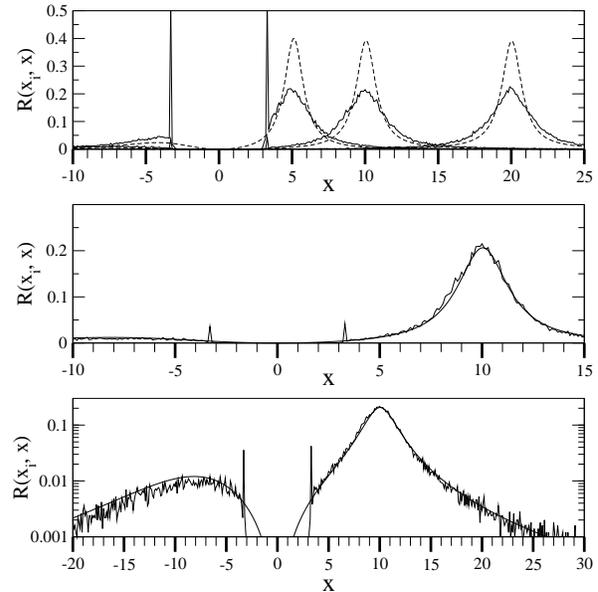} }
\caption{{\bf Dust-free HI Frequency Redistribution} {\sl Top:} Reflection-line profiles for $x_i=5, \ 10, \ 20$ are shown.  The solid lines are simulations, while the dashed lines are the analytic predictions $R_{\rm anlyt}(x)$, eqn. (\ref{R_neuf}). There are significant discrepancies, as discussed in the text. {\sl Middle}:  The reflection line profile for $x_i=10$ is shown.  
The jagged solid line is the simulation, while the smooth solid line is the fit $R(x_i,x; \alpha)$, eqn. (\ref{R_fit}), with $\alpha=45/2$ and ${\tilde x}_i$ given by eqn. (\ref{xi_shift}). This simple rescaling of the analytic redistribution function gives accurate fits to the simulations. {\sl Bottom}:  The same as the middle panel, but shown on a logarithmic scale and over a larger frequency range.  In all three plots, the sharp peaks at $x\approx\pm 3$ are an artifact of the acceleration scheme that we use;  we do not calculate any core scatters, and photons which escape the core are all placed at $x=\pm 3.3$.  Exact simulations indicate that there is indeed a ``pile up'' of photons just outside the core, but the peaks are not as sharp.  In any event, the number of photons in the peaks are relatively insignificant.}
\label{figure_R_HI}
\end{center}
\vspace{0cm}
\end{figure}
\begin{figure}
\begin{center}
\vspace{0.5cm}
\scalebox{0.42}[0.42]{
\includegraphics[angle=0]{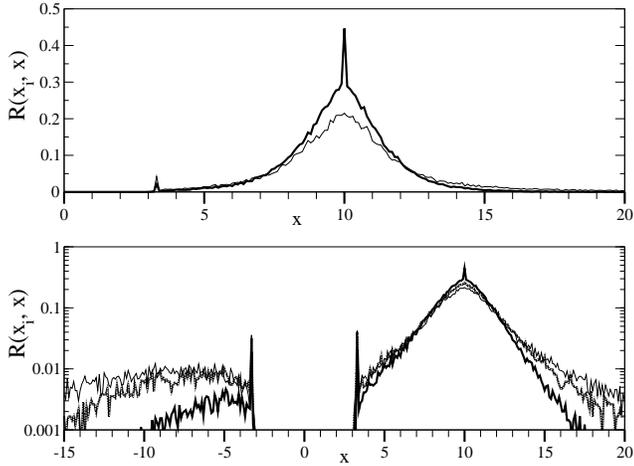} }
\caption{{\bf Dusty HI Frequency Redistribution}  {\sl Top}:  The simulated reflection line profiles for $x_i=10$ and three different $\sigma^a_{-21}$ are shown: the thin line is $\sigma^a_{-21}=0$, the light thick line is $\sigma^a_{-21}=1$ and the dark thick line is $\sigma^a_{-21}=10$.  {\sl Bottom}:  The same as the top panel but on a logarithmic scale and over a larger frequency range. 
Over the frequency range $0 < x < 15$, the frequency redistribution function $R(x,x_{i})$ differs by $ < 20 \%$ between $0 < \sigma_{-21}^{a} < 1$; hence, to lowest order it is independent of dust content for the cases we are interested in.
}
\label{figure_R_dusty}
\end{center}
\vspace{0cm}
\end{figure}
When compared to simulations, $R_{\rm anlyt}(x)$ is inaccurate in two respects.  First, the actual peaks are shifted by an amount $-2/x_i$ (towards line center) compared to $R_{\rm anlyt}$.  This can be compensated for by using a shifted incident frequency ${\tilde x}_i$ in place of $x_i$, where
\begin{equation}\label{xi_shift}
{{\tilde x}_i}\equiv x_i-2/x_i\ .
\end{equation}
Second, the peaks are significantly flatter than those given by $R_{\rm anlyt}$.  It is not surprising that $R_{\rm anlyt}$ is inaccurate, since the analytic results of \cite{neufeld90} only hold when the photons traverse a large line center optical depth, $a\tau_0\geqsim10^3$.  The bulk of photons that reflect off a surface do not traverse such a large optical distance, and so the analytic result may not be accurate, which seems to be the case.  However, 
one can obtain an excellent approximation to the redistribution function by modifying eqn. (\ref{R_neuf}) to include an additional fitting parameter $\alpha$ (see Appendix \ref{appendix_Redistribution} for details): 
\begin{equation}\label{R_fit}
R(x_i,x;\alpha)=\frac{3\sqrt{\alpha}}{\pi}\frac{x^2{{\tilde x}_i}^2}{ \alpha {{\tilde x}_i}^4+\left(x^3-{{\tilde x}_i}^3\right)^2 }\ .
\end{equation}
Note that $R(x_i, x; \alpha)$ is guaranteed to be normalized for all $\alpha$.  For example, $R_{\rm anlyt}$ is recovered with $\alpha=6$ and ${\tilde x}_i=x_i$.  We find that our simulations for pure HI are well fit by $\alpha=45/2$, with ${\tilde x}_i$ given by eqn. (\ref{xi_shift}).  To generate random frequencies $x$ that obey the probability distribution $R(x_i,x;\alpha)$, one draws a random univariate $u\in[0,1]$, and sets
\begin{equation}
u=\int_{-\infty}^x{\rm d}x' R(\tilde{x}_i, x'; \alpha)\equiv F(x)\ .
\end{equation}
The frequency $x$ is then given by functional inversion, $x=F^{-1}(u)$.  Carrying out these steps on eqn. (\ref{R_fit}), done in Appendix \ref{appendix_Redistribution}, we find that the exiting frequencies $x$ which obey the probability distribution $R(x; \alpha)$ can be generated by the equation
\begin{equation}\label{exiting_x}
x=\left[{{\tilde x}_i}^3-{{\tilde x}_i}^2\sqrt{\alpha}\, \tan\left(\pi u\right)\right]^{1/3}
\end{equation}
where $u\in[0,1]$ is a random univariate.  

When dust is included, the profile peaks become slightly sharper and the tails fall off slightly faster.  However, as shown in Figure \ref{figure_R_dusty}, the pure HI distribution closely matches the dusty distribution when $\sigma^a_{-21}<10$. In practice, we shall therefore always adopt eqn. (\ref{R_fit}) with $\alpha=45/2$ for frequency redistribution in the line wing, since it is accurate except for galaxies with highly supersolar metallicities, which is unlikely at high redshift. 

\begin{figure}
\begin{center}
\vspace{0.5cm}
\scalebox{0.4}[0.4]{
\includegraphics[angle=0]{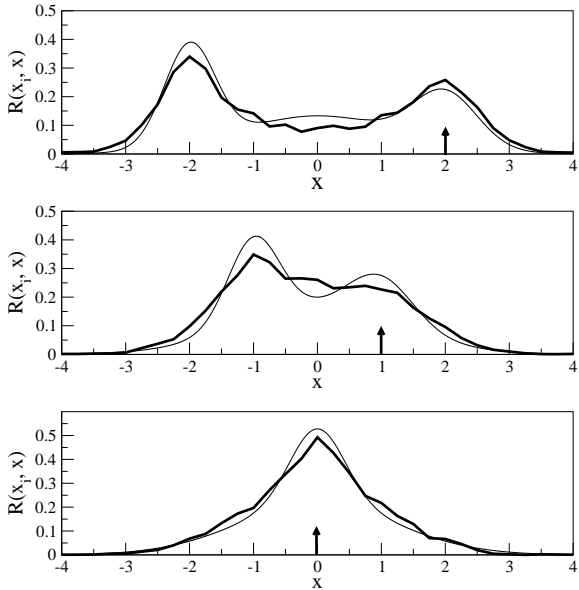} }
\caption{{\bf Dust-free Core Redistribution}  The simulated emission line profiles are shown for three different incident frequencies that lie within the Doppler core.  As indicated by the arrows, the panels from top to bottome correspond to $x_i=2, \ 1, \ 0$.  In each panel, the thick line is the exact simulation data and the thin line is the fitting formula eqn. (\ref{Rcore_fit}). }
\label{fig_R_core}
\end{center}
\vspace{0cm}
\end{figure}

When the incident frequency is in the Doppler core, the analytic fit, eqn. (\ref{R_fit}), breaks down, and the emission profile takes on an entirely different form.  The emission profile roughly breaks down into two principal components: photons that escape after only a few scatters, and photons that scatter enough times that they reach line center before escape.  The former photons retain some ``memory'' of their incident frequency $x_i$, and produce emission peaks at $x=x_i$ and $x=-x_i$.  (the peak at $-x_i$ is from photons that have undergone a single particle back-scattering off atoms with velocity $V=x_i V^{\rm dop}$ along the photon propagation direction). The latter photons lose all memory of their initial frequency, and produce a broader emission peak centered on $x=0$.  Accordingly, we fit the core redistribution function $R^{\rm core}(x_i,x)$ with three gaussians, centered on $-x_i, \ 0,$ and $x_i$, respectively.  Let us define the gaussian distribution with r.m.s. frequency $\sigma$:
\begin{equation}
G(x,\sigma)\equiv \frac{1}{\sigma\sqrt{2\pi}} \ e^{-\frac{x^2}{2\sigma^2}}\ .
\end{equation}
By comparing to exact simulations, shown in Figure \ref{fig_R_core}, we find that a decent fit is given by
\begin{equation}\label{Rcore_fit}
\begin{array}{l}
R^{\rm core}(x_i, x)=\\ 		
\quad\quad[1-P(x_i)]\left[\frac{3}{5}G(x+x_i, A)+\frac{2}{5}G(x-x_i, B)\right]  \\
\quad\quad{}+ P(x_i)G(x,C ) \ ,
\end{array}
\end{equation}
where 
\begin{eqnarray}
P(x_i)&=&0.7e^{-0.2 x_i}\\
A&=&0.4 \nonumber \\
B&=&0.5 \nonumber \\
C&=&1.25 \nonumber
\end{eqnarray}
The above fit works well when dust is included, since the effect of dust on photons in the Doppler core is small in absolute terms (see \S \ref{section_acceleration_scheme}).

In summary, a simple analytic prescription for the surface frequency redistribution function $R(x_i,x)$ is to use equation (\ref{R_fit}) with $\alpha=22.5$ and $\tilde{x}_i=x_i-2/x_i$ when the incident photon is in the line wing $|x_i|\geq 3$, while equation (\ref{Rcore_fit}) can be used when the incident photon is in the Doppler core $|x_i|<3$. 

\subsection{Surface Kinematics}\label{section_kinematics}
In this section we discuss two kinematic effects of surface scattering.  First, we calculate the average scattering angle cosine $g=\langle \cos\theta_{scat}\rangle$, where $\theta_{\rm scat}$ is the angle between the incident and exiting direction.  Second, we calculate the net frequency shift $\Delta V$ due to the Doppler shift induced by scattering off a moving surface.  In each case, we consider isotropic and perpendicularly incident photons, 
and use the surface scattering angular distribution $D_{ss}(\theta)$, eqn. (\ref{exit_angle}).
The case of perpendicularly incident photons is of interest because it applies to photons bouncing around inside a spherical shell;  most photons that strike a point on the surface last reflected off the far side of the shell, and hence incident photons have a strong bias to lay along the surface's perpendicular.

We use the following conventions: the outward normal to the surface $\hat{n}_s$ defines the $\hat{z}$ direction, an incident photon has direction $\hat{n}_1$ such that $\hat{n}_s\cdot\hat{n}_1<0$ with polar angle $\theta_1$ and azimuthal angle $\phi_1$, an exiting photon has direction $\hat{n}_2$ such that $\hat{n}_s\cdot\hat{n}_2>0$ with polar and azimuthal angles $\theta_2$ and $\phi_2$.  For example, in Cartesian co-ordinates, $\hat{n}_1=(\cos\phi_1 \sin\theta_1, \ \sin\phi_1\sin\theta_1, \ \cos\theta_1)$.
The surface has a bulk velocity $\vec{V}_s$, with a perpendicular component $V_{\perp}\equiv \hat{n}_s\cdot \vec{V}_s$. 
For an isotropic angular distribution of incident photons, the polar angle distribution of incident photons is $D_{iso}(\theta_1)=\sin\theta_1$, 
which is normalized to unity over $\theta_1\in[\pi/2, \pi]$, and $\phi_2$ is uniformly distributed over $2\pi$.  For exiting photons, eqn. (\ref{exit_angle}), the polar angle distribution is given by $D_{ss}(\theta_2)=2\sin\theta_2\cos\theta_2$, which is normalized to unity over $\theta_2\in[0, \pi/2]$, and $\phi_2$ is uniformly distributed over $2\pi$.

\subsubsection{The Average Scattering Angle}
The scattering angle cosine, also called the scattering asymmetry parameter, is defined by $g\equiv \langle \cos\theta_{\rm scat}\rangle=\langle \hat{n}_1\cdot \hat{n}_2\rangle$.  For an isotropic incident angle distribution, the average over the angles is
\begin{eqnarray}
g_{\rm iso}&=&  \int_{\pi/2}^\pi \dd\theta_1\int_ 0^{2\pi}\dd \phi_1  \int_{0}^{\pi/2} \dd\theta_2\int_ 0^{2\pi}\dd \phi_2 \ \ \times \\
  &&\big\{ D_{iso}(\theta_1) D_{ss}(\theta_2) \  \hat{n}_1\cdot \hat{n}_2 \big\} \nonumber \ .
\end{eqnarray}
Azimuthal symmetry eliminates all the terms in $\hat{n}_1\cdot\hat{n}_2$ that have an $\hat{x}$ and $\hat{y}$ contribution, leaving just the term $\cos\theta_1 \cos\theta_2$.  The integral over $\theta_1$ and $\theta_2$ separate, giving
\begin{equation}\label{eqn:g_iso}
g_{\rm iso}=-\frac{1}{3}\ .
\end{equation}  
The same steps can be carried out for perpendicularly incident photons, $\hat{n}_1=-\hat{z}$, resulting in
\begin{equation}
g_{\perp}=-\frac{2}{3}\ .
\end{equation}
Surface scattering is characterized by a net average back-scatter, $g<0$.

\subsubsection{Frequency Shift Due to a Bulk Surface Velocity}\label{section_KinematicsVelocity}
If the surface has a bulk velocity, then the frequency of the scattered photons suffer a net Doppler shift, due purely to the surface motion.  Consider a photon with frequency $V$ striking a moving surface. In the surface rest frame, the photon has incident frequency $V'=V-\hat{n}_1\cdot \vec{V}_s$.  An exiting photon with frequency $V''$ in the surface rest frame has an exiting frequency $V'''=V''+\hat{n}_2\cdot\vec{V}_s$ in the original (``lab'') frame.  Therefore the surface motion induces a net frequency shift
\begin{equation}
\Delta V_{\rm sm}=\left(\hat{n}_1-\hat{n}_2\right)\cdot \vec{V}_s\ ,
\end{equation}
which is in addition to any frequency shift from scattering within the surface.  Averaging over an isotropic incident angle gives 
\begin{equation}
\langle \hat{n}_1\rangle_{\rm iso}=-\frac{1}{2} \hat{z}\ .
\end{equation}
For perpendicularly incident photons, we simply have $\langle \hat{n}_1\rangle=-\hat{z}$.  For the exiting direction, averaging over $D_{ss}(\theta_2)$ gives
\begin{equation}
\langle \hat{n}_2\rangle= \frac{2}{3}\hat{z}\ .
\end{equation}
Thus, the average frequency shift for isotropic and perpendicularly incident photons are, respectively, 
\begin{eqnarray}
\langle \Delta V_{\rm sm} \rangle_{\rm iso}&=& \frac{7}{6}V_{\perp}\\
\langle \Delta V_{\rm sm} \rangle_{\perp}&=& \frac{5}{3}V_{\perp}\ ,
\end{eqnarray}
where, as stated above, we define $V_{\perp}\equiv \hat{n}_s\cdot \vec{V}_s=\hat{z}\cdot\vec{V}_s$.  If the surface is moving away from (towards) the incident photons, $V_{\perp}<0$, then surface scattering causes a net redshift (blueshift) of the photon frequency.  

\section{Analytic Multi-phase Transfer}\label{section_mp}
In this section, we build an analytic model for estimating the escape fraction and line width for \Lya escaping from a multi-phase region composed of dusty, optically thick clumps. Compared against Monte-Carlo simulations, these analytic estimates give remarkably accurate results. We consider both stationary clumps and clumps with a Maxwellian bulk velocity distribution, but postpone the discussion of bulk gas outflow/inflow until \S
\ref{section_outflow}.  We derive a general photon escape fraction formula in terms of two parameters: the mean number of surface scatters in the absence of absorption $\N_0$, and the cloud albedo $\epsilon_c$, for the case of coherent scattering. The parameter $\N_0$ is purely geometric and independent of photon frequency, as long as clouds are very optically thick. We derive fits for $\N_0$ for a variety of multi-phase geometries of astrophysical interest, and compare the resulting analytic escape fractions to Monte-Carlo simulations of coherent scattering.  

Applying the coherent scattering formulas to \Lya radiative transfer requires some care, since in this case $\epsilon_c$ is frequency dependent through the frequency dependence of the Ly$\alpha$ scattering cross-section. The cloud albedo, therefore, changes as the photon executes a random walk in frequency. Nonetheless, we find that the analytic formula for coherent scattering can be extended to \Lya scattering as long as $\epsilon_c$ is evaluated at the characteristic escape frequency. In \S \ref{section_line_widths}, we study how the characteristic escape frequency depends upon the frequency redistribution per surface scatter and the random bulk motion of the clumps. We derive estimates of the escaping line profile r.m.s. width and FWHM, and compare these to simulations of repeated surface scatterings.  

\subsection{Escape Fractions}\label{section_analytic_coherent}
A more detailed prediction for the escape fraction can be given than the simple scaling laws in \S \ref{section_scalings}.  We derive a generic escape fraction formula in terms of the probability of absorption per scattering, $\epsilon$, and the average number of scatters for escape in the absence of absorption, $\N_0$.  In the context of multi-phase transfer, each ``scatter'' refers to an entire surface scattering, in which case $\epsilon$ is the cloud albedo: $\epsilon\equiv \epsilon_c$.  

Central to the analysis is the average number of interactions for escaping photons, $\N$.  Let $\epsilon$ be the average absorption probability per interaction, and define $D(n)$ to be the probability distribution for photons to interact $n$ times before escaping, when $\epsilon=0$ (i.e., in the absence of absorption). For a constant $\epsilon$, the escape fraction can be written
\begin{equation}
\fesc=\sum_{n=0}^{\infty}(1-\epsilon)^nD(n)\ .
\end{equation}
The average number of interaction for escaping photons can likewise be written as
\begin{equation}
\N=\frac{1}{\fesc}\sum_{n=0}^{\infty}(1-\epsilon)^n\,nD(n)\ .
\end{equation}
From these two expressions, we derive
\begin{equation}\label{Nesc}
\N=-(1-\epsilon)\frac{\dd}{\dd\epsilon} \ln \fesc\ .
\end{equation}
This can be inverted to express $\fesc$ in terms of $\N$:
\begin{equation}
\fesc=\exp\left(-\int_0^\epsilon \dd \epsilon' \frac{\N(\epsilon')}{1-\epsilon'}\right) \ .
\end{equation}
From eqn. (\ref{Nesc}), $\N_0\equiv \N(0)$ is given by
\begin{equation}\label{N0}
\N_0=\lim_{\epsilon\rightarrow 0}\; -\frac{\dd}{\dd\epsilon} \ln \fesc\ .
\end{equation}

To derive escape fractions in an arbitrary geometry, let us consider two limits: narrowly beamed forward scattering, and front-back symmetric scattering. The latter refers to the $g\equiv \langle \cos\theta_{\rm sct}\rangle =0$ case, where there is equal probability to scatter in the forward and backward directions; both isotropic and dipole scattering satisfy this, though dust grains ($g_d=0.5$) and cloud surfaces ($g_{\rm iso}=-1/3$, from eqn. (\ref{eqn:g_iso})), do not.  As we shall show, however, the escape fraction for scattering for any value of $g< 1$ can be based upon the $g=0$ case as long as there are sufficient scatters that the photon's trajectory is randomized.

The case of purely forward scattering is trivial: since the photon does not change direction, we have
\begin{equation}\label{fesc_g=1_N0}
f_{\rm e} = e^{-\tau_{\rm abs}} = e^{-\epsilon \tau} = e^{-\epsilon \N_{0}},
\end{equation}
where we have used eqn. (\ref{N0}) in the final step. Note that the escape fraction is same as if there were no scatterers, since scattering does not alter the photon trajectory.
For front-back symmetric scattering ($g=0$), the escape fraction is (see Figure \ref{figure_1Dfesc} in Appendix \ref{section_1Dg}):
\begin{equation}\label{fesc_g=0_N0}
\fesc=\frac{1}{\cosh(\epsilon^{1/2}\sqrt{\tau^2+2\tau})}= \frac{1}{\cosh\left(\sqrt{2\epsilon \N_0}\right)} \ .
\end{equation}
where we applied eqn. (\ref{N0}) to obtain $\N_0=\frac{1}{2}(\tau^2+2\tau)$ in the second step\footnote{Within the Eddington approximation, the escape fraction from a source in the mid-plane of a slab is $\fesc=1/\cosh\left(\sqrt{3} \ \epsilon^{1/2} \tau \right)$, as derived, for example, in \citet{neufeld91}.  Instead, we propose eqn. (\ref{fesc_g=0_N0}), which from our simulations is more accurate for 1-D scattering (see Appendix \ref{section_1Dg}).}. Although this formula for the escape fraction is derived assuming $g=0$, we show in Appendix \ref{section_1Dg} that this formula is valid for any $g$ as long as there are enough scatters to randomize the photon's direction. Specifically, if $\N_0 > n^{*}(g)$ (where $n^{*}(g)$ is given by eqn. (\ref{nstar_g})), then eqn. (\ref{fesc_g=0_N0}) can still be used. Otherwise, eqn. (\ref{fesc_g=1_N0}) is a better approximation to the escape fraction---since a situation with such few scatters and/or a strongly peaked forward scattering profile converges to the straight-line trajectory case.
Note that in both these limits, the escape fraction only depends upon a single parameter, $\epsilon\N_0$.  The average number of interactions required for a photon to be absorbed is $\N_{a}=1/\epsilon$,
so the controlling parameter is equivalent to $\epsilon\N_0 = \N_0/\N_a$.

The average number of interactions for an escaping photon, $\N$, can be derived by applying eqn. (\ref{Nesc}) to the appropriate escape fraction formulae,  eqn. (\ref{fesc_g=1_N0}) or  eqn. (\ref{fesc_g=0_N0}).  For straight-line trajectories, the result is
\begin{equation}
\N=(1-\epsilon)\N_0\ ,
\end{equation}
while for the random walk trajectories
\begin{equation}\label{Nesc_N0}
\N=(1-\epsilon)\N_0 \, \frac{\tanh\left(\sqrt{2\epsilon \N_0}\right)}{\sqrt{2\epsilon \N_0}}\ .
\end{equation}
In the limit $\sqrt{\epsilon\N_0}\ll 1$, we find $\N\rightarrow \N_0-\epsilon\N_0(\frac{2}{3}\N_0+1)$, and in the limit $\sqrt{\epsilon\N_0}\gg 1$, we have $\N\rightarrow \frac{(1-\epsilon)}{2\epsilon}\sqrt{2\epsilon\N_0}$.

The entire effect of the cloud geometry is characterized by a single number, $\N_0$.  Directly calculating $\N_0$ from a given geometry is not practical except for the simplest geometries.  In general, given a clump geometry, $\N_0$ must be computed via a simulation, where one can use the exiting angle distribution, eqn. (\ref{exit_angle}), for each surface reflection. 
In practice, we find that for many generic geometries expected to crop up in astrophysical applications, an appropriate ``line of sight''-averaged $\N_0$ can be accurately determined from simple geometric parameters (such as the cloud covering factor $f_{C}$). We now proceed to do so.

\subsection{Example Geometries}\label{section_geometries}
We now test the accuracy of the analytic formula eqn. (\ref{fesc_g=0_N0}), and investigate its application in a variety of multi-phase geometries.  For each geometry, we fit $\N_0$ as a function of the appropriate natural geometric parameter (such as cloud covering factor $f_{C}$) using Monte Carlo simulations.  We then calculate the escape fraction using $\N_0$ and eqn. (\ref{fesc_g=0_N0}), and compare this to simulations.  In several cases, we find that we obtain better fits by rescaling with an order unity fitting parameter $\kappa$, where we use $\kappa\epsilon_c\N_0$ in place of $\epsilon_c\N_0$ in the escape fraction formulae eqn. (\ref{fesc_g=0_N0}). In general, even with no correction factor, eqn. (\ref{fesc_g=0_N0}) is accurate to $\leqsim 10\%$ when $\fesc\geq 10\%$, and is generally correct to within a factor $\leqsim 2$ for $\fesc<10\%$.  When the escape fraction is very small, the photons that {\sl do} escape comprise the rare trajectories.  As their transfer behavior can depend sensitively on the specifics of the geometry, it is not surprising that eqn. (\ref{fesc_g=0_N0}) breaks down when $\fesc$ is very small. In any case, \Lya emission is undetectable for such small $\fesc$, so these cases are of little observational consequence.

Some notes about our Monte-Carlo simulations: for simplicity, we assume that the region in between the scattering surfaces is empty (we justify this assumption in \S \ref{section_ICM_gas}). We also assume that scattering surfaces are extremely optically thick, so that the surface scattering approximations of \S \ref{section_surfaces} apply.  In particular, when a photon hits a gas surface, it has a net probability $\epsilon_c$ of being absorbed;  if it survives, then its exiting angle relative to the surface normal follows the distribution $D_{ss}(\theta)=\sin2\theta$, and the exit location is the same as the point of incidence.

\subsubsection{Spherical Clumps}\label{geometry_spheres}
\begin{figure}
\begin{center}
\vspace{0.5cm}
\scalebox{0.5}[0.5]{
\includegraphics[angle=0]{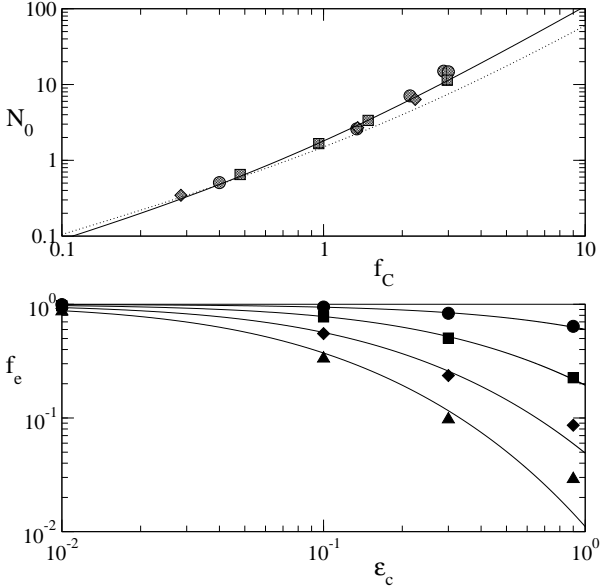} }
\caption{{\bf Random Spheres Geometry}
Spherical clouds of radius $r$ randomly populate a spherical region with radius $R$.  The circles have clouds with the same radius, $r/R=1/20$.  {\sl Top}:
The number of scatterings $\N_0$ as a function of the cloud covering factor $f_{C}$.
The squares have clouds with the same radius $r/R=1/40$.  The diamonds have a distribution of cloud sizes with the distribution $D(r)\propto 1/r^2$, with a minimum cloud radius $r/R=3/200$ and a maximum cloud radius $r/R=1/20$.  The solid line is the fit eqn. (\ref{N0_spheres}), while the dotted line is the naive 1-D relation $\N_0=\frac{1}{2}({f_C}^2+2f_C)$, extrapolated from $\N_0=\frac{1}{2}( \tau^{2}+2\tau)$, and $\tau \rightarrow f_{C}$ (in fact, $\tau \rightarrow \sqrt{2} f_{C}$ is the appropriate substitution). {\sl Bottom}: Photon escape fraction $f_{\rm e}$ as a function of cloud albedo $\epsilon_c$.
From top to bottom, the covering factors are $f_C=0.48, 1.34, 2.14, 3.0$.  The lines are based on eqn. (\ref{N0_spheres}). }
\label{figure_spheres}
\end{center}
\vspace{0cm}
\end{figure}

The canonical example of a multi-phase geometry is a spherical region populated by randomly placed, optically thick spherical clumps. Such clumps tend to be cool-phase gas such as molecular clouds, which arose via thermal instability. The natural geometric parameter is the mean number of clouds intersected along a random line of sight, called the cloud covering factor $f_C$.  The covering factor is analogous to the interaction optical depth $\tau$ for homogenous media.  For a central source, $f_C$ is measured from the region center to the edge.  We computed $\N_0$ for various covering factors and clump radii distributions.  As shown by the solid line in the top panel of Figure \ref{figure_spheres}, we find that a general fit for {\it any} clump radii distribution is
\begin{equation}\label{N0_spheres}
\N_0={f_C}^2+\frac{4}{5}f_C\ .   
\end{equation}
This highlights the fact that when surface scattering applies, the spherical clump geometry does not depend upon the size distribution of the clumps nor their volume filling fraction;  the entire geometry is characterized by a single parameter, $f_C$. This was postulated by \citet{neufeld91}; we have confirmed this insight in our simulations. The scalings in eqn. (\ref{N0_spheres}) can be compared against the usual random walk formulae, $\N \approx \tau$ (for $\tau \ll 1$), and $\N \approx \tau^{2}$ (for $\tau \gg 1$), for scattering with front-back symmetry $g=0$ (e.g., Rybicki \& Lightman 1979, p.35); here, $f_C$ plays the role of $\tau$. Our scalings are slightly different, since $g=-1/3$ for our clouds.
For 1-D scattering, $\N_0=\frac{1}{2}(\tau^2+2\tau)$ (shown as a dotted line in Fig. \ref{figure_spheres}).
By comparison against eqn. (\ref{N0_spheres}), the spherical clump model is analogous to 1-D radiative transfer with the substitution $\tau\approx \sqrt{2} f_C$.
In the bottom panel of Figure \ref{figure_spheres} we show that the escape fraction formula, eqn. (\ref{fesc_g=0_N0}), works well for a variety of covering factors for constant $\epsilon_c$.

While the covering fraction is an unknown free parameter, it is easy to see why $f_{C} \sim 1$ is reasonable. Suppose the cold clouds constitute a mass fraction $f_{M}$ of the galaxy, and are overdense by a factor $\delta$ relative to the intercloud medium. Assuming pressure balance, $\delta \sim T_{\rm ICM}/T_{c} \sim 100$, for $T_{c} \sim 10^{4}$K and $T_{\rm ICM} \sim 10^{6}$K for the cloud and inter-cloud medium temperatures respectively. The volume filling factor of clouds is then $f_{V}=f_{M}/\delta$. The number density of clouds is $n_{c} \sim f_{V}/V_{c}$, where $V_{c} \sim r_{c}^{3}$ is the volume of a typical cloud. The cloud covering factor is:
\begin{equation}
f_C \sim n_{c} \sigma_{c} r_{\rm gal} \sim f_{V} \frac{r_{\rm gal}}{r_{c}} \sim 3 \left( \frac{f_{M}}{0.3} \right) \left( \frac{\delta}{100} \right) \left( \frac{r_{\rm gal}/r_{c}}{10^{3}} \right) 
\end{equation}

\subsubsection{Random Surfaces}\label{section_r-surfaces}
\begin{figure}
\begin{center}
\vspace{0.5cm}
\scalebox{0.5}[0.5]{
\includegraphics[angle=0]{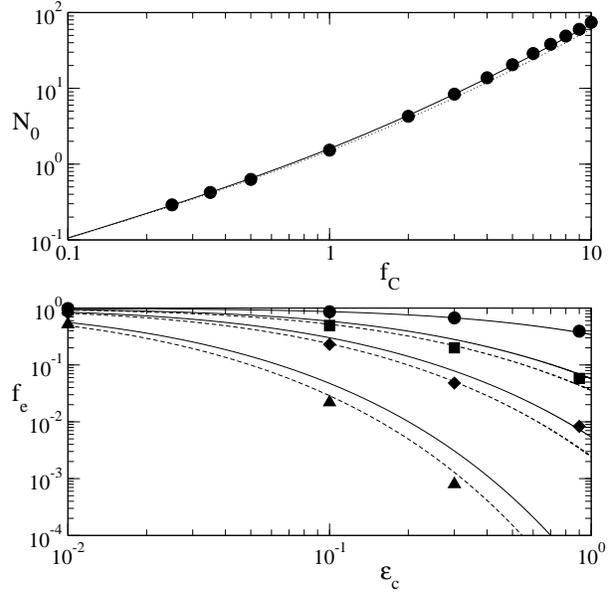} }
\caption{{\bf Random Surfaces Geometry}  {\sl Top}:  $N_0$ vs. $f_{C}$. Points are simulation results, while the solid line is the fit eqn. (\ref{N0_rs}).  As in the top panel of Figure \ref{figure_spheres}, the dotted line is the naive 1-D power law $\N_0=\frac{1}{2}({f_C}^2+2f_C)$.
{\sl Bottom}: $f_{\rm e}$ vs. $\epsilon_{c}$. From top to bottom, the covering factors are $f_C=1, \, 3,\, 5,\,10$.  The solid lines use $\epsilon_c\N_0$ in the escape fraction formula eqn. (\ref{fesc_g=0_N0}), while the dashed line uses $\kappa \epsilon_c\N_0$, with a correction factor $\kappa=1.29$.}
\label{figure_rs}
\end{center}
\vspace{0cm}
\end{figure}

An abstraction of the spherical clump geometry is to have the photons strike a surface after traveling a distance $\ell$, where $\ell$ is drawn from a given probability distribution, and where each surface is randomly oriented with respect to the photon direction.  We have investigated the exponential distribution $\exp(-f_C\ell/R)$, where $R$ is the radius of the region and $f_C$ is the covering factor.  The photon escapes once it leaves the spherical region of radius $R$.  When a photon traveling in the direction $\hat{n}_p$ strikes a surface, the outward normal of the surface $\hat{n}_s$ is drawn from an isotropic distribution such that $\hat{n}_p\cdot\hat{n}_s<0$.  As shown by the top panel in Figure \ref{figure_rs}, a good fit for the scattering number is
\begin{equation}\label{N0_rs}
\N_0=\frac{3}{5}{f_C}^2+f_C
\end{equation}
The random surfaces model is faster to simulate, and any results reliably apply to the spherical clump model when expressed in terms of $\N_0$.  The bottom panel of Figure \ref{figure_rs} compares the escape fraction formula eqn. (\ref{fesc_g=0_N0}) to simulations.

This procedure for generating random surfaces can in fact be used to quickly simulate any arbitrary geometry, with a suitable characterization of the probability distribution of path lengths $\ell$ and surface orientations $\hat{n}_s$, though, of course, the fitting formulae for $\N_0$ will depend on these quantities.

\subsubsection{Shell with Holes}\label{section_shell_holes}
\begin{figure}
\begin{center}
\vspace{0.5cm}
\scalebox{0.5}[0.5]{
\includegraphics[angle=0]{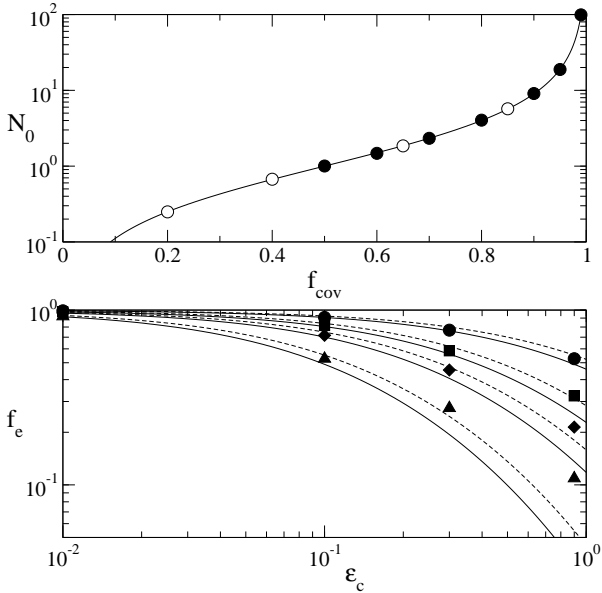} }
\caption{{\bf Shell with Holes Geometry}  {\sl Top}: The scattering number $\N_0$ as a function of the shell covering factor $f_{\rm cov}$. Simulations were run for shells with one gap (light circles) and five gaps (dark circles), which fall almost exactly on the fit eqn. (\ref{N0_shell}).  As expected, the scattering number $\N_0$ only depends on $f_{\rm cov}$, and not the number of gaps. {\sl Bottom}: $f_{\rm e}$ vs. $\epsilon_{c}$. From top to bottom, the covering factors are $f_{\rm cov}=0.5, \, 0.7,\, 0.8, \,0.9$.  The lines show the fitting formula eqn. (\ref{fesc_g=0_N0}), with $\kappa \epsilon_c\N_0$ as the variable, with $\kappa=1$ (solid lines), and $\kappa=0.8$ (dashed lines).}
\label{figure_shell}
\end{center}
\vspace{0cm}
\end{figure}

Another geometry that frequently arises in astrophysics is that of a shell of material surrounding a photon source---for instance, in stellar and galactic outflows. Much work has been done on \Lya radiative transfer through opaque shells \citep{Tenorio-Tagle99, ahn03, ahn04},
but the effect of gaps in the shell has not been investigated.  Since a completely homogeneous shell of gas is rarely, if ever, observed, a shell with holes is an interesting geometry.  The natural geometric parameter here is the fraction of the solid angle covered by the shell, $f_{\rm cov}$ (i.e., the gaps comprise a total solid angle $4\pi(1-f_{\rm cov})$).  To estimate $\N_0$, assume that during each bounce the photon has a probability of $(1-f_{\rm cov})$ to escape through a gap.  This leads to the expression
\begin{equation}\label{N0_shell}
\N_0\approx f_{\rm cov}/(1-f_{\rm cov})\ ,
\end{equation}
which is shown by the solid line in the top panel of  Figure \ref{figure_shell}.  We computed $\N_0$ for shells with a random placement of non-overlapping circular gaps, and found that many small gaps were equivalent to fewer large gaps with the same $f_{\rm cov}$, as to be expected.  As shown in the bottom panel of Figure \ref{figure_shell}, the analytic form for $\fesc$ does well where expected, and only breaks down when $f_{\rm cov}$ is near unity and $\epsilon_{c}\geqsim 0.3$.
This agreement is quite remarkable, given that the escape fraction formula (\ref{fesc_g=0_N0}) was derived assuming a very different scattering particle geometry (homogeneous media). This gives us confidence that the impact of geometry on the escape fraction can indeed be encapsulated by a single parameter $\epsilon_{c} \N_0$.

\subsubsection{Open Ended Tube}
\begin{figure}
\begin{center}
\vspace{0.5cm}
\scalebox{0.5}[0.5]{
\includegraphics[angle=0]{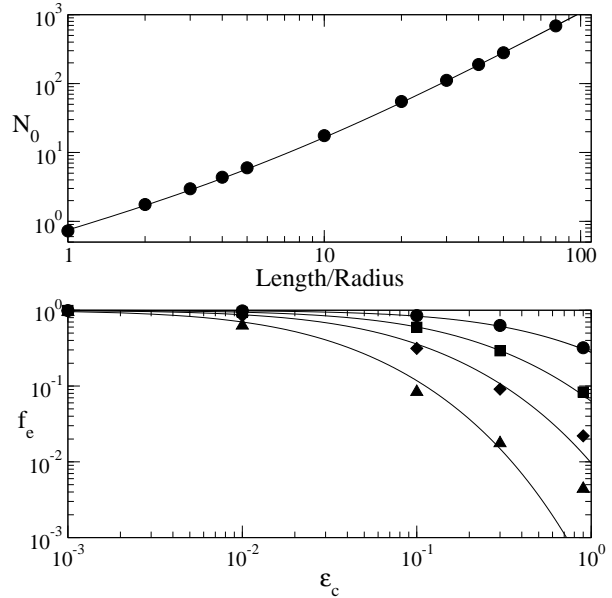} }
\caption{{\bf Tube Geometry} {\sl Top}:  The solid line is the fit for $\N_0$, eqn. (\ref{N0_tube}), and the circles are simulation results.  {\sl Bottom}:  From top to bottom, the length to radius ratios are  $L/R=2, \, 5, \, 10, \, 20$.  The solid lines are given by eqn. (\ref{fesc_g=0_N0}), with no correction factor ($\kappa=1$).}
\label{figure_tube}
\end{center}
\vspace{0cm}
\end{figure}

To illustrate how well photons can escape through cracks and gaps in optically thick material, we consider photons escaping from the middle of an open-ended tube.  This geometry may apply to star forming regions or AGNs where outflows have punched (bipolar) cavities through a surrounding gas cloud, allowing the photons to escape through the cavities (e.g. \citet{Shopbell98}), although we do not investigate the dependence on the opening angle.  The natural geometric parameter is the ratio $L/R$, where $L$ is the total length of the tube and $R$ is the tube's radius.  The average length traveled along the tube per scatter is $\ell\sim R$.  Therefore $\N_0$ is estimated from $L/2\approx \sqrt{\N_0} \ell$, which implies $\N_0\sim (L/R)^2$.  As shown by the solid line in the top panel of Figure \ref{figure_tube}, a decent fit for the number of surface scatters is
\begin{equation}\label{N0_tube}
\N_0=\frac{1}{10}(L/R)^2+\frac{13}{20}(L/R)
\end{equation}
As shown in the bottom panel of Figure \ref{figure_tube}, the escape fractions fit the data very well, except as $\epsilon_{c} \rightarrow 1$. As previously discussed, for such low escape fractions, rare escaping photons follow unusual trajectories not well captured by our formalism. In any case, for such low escape fractions, \Lya emission cannot be observed, and the results have no observational relevance.

\subsection{Line Widths}\label{section_line_widths}
In this section we shall consider two separate effects on the \Lya frequency as it escapes:  the frequency redistribution due to thermal motion of atoms, as well as random bulk motion of the scattering surfaces.  The effect of global outflow/inflow on the line profile is discussed separately, in \S \ref{section_outflow}.  We consider two simple diagnostics of the profile: the r.m.s. frequency shift $\Sigma$ and the FWHM $\Gamma$.  We ran simulations of dust-free surface scattering to determine $\Sigma$ and $\Gamma$ as a function of the number of surface scatterings, $n_{ss}$.  Although escaping photons vary in the number of scattering surfaces they encounter before escape, in practice we find that line profiles can be accurately characterized by the {\sl average} number of scattering surfaces encountered.  We now derive formulas for $\Sigma(n_{ss})$ and $\Gamma(n_{ss})$ for resonant scattering frequency redistribution and random bulk surface velocities.  
\begin{figure}
\begin{center}
\vspace{0.5cm}
\scalebox{0.42}[0.42]{
\includegraphics[angle=0]{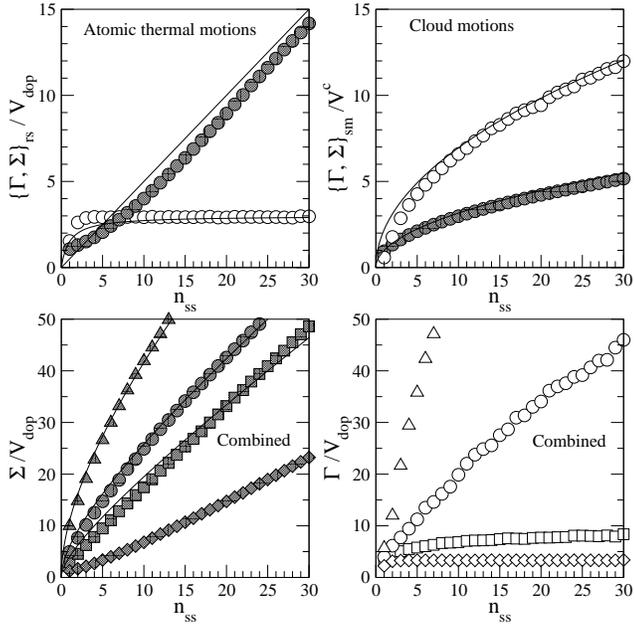} }
\caption{{\bf Line Widths}  {\sl Top Left}:  The r.m.s frequency shift $\Sigma$ (filled circles) and the FWHM $\Gamma$ (open circles) are shown as a function of the $n^{th}$ surface scatter $n_{ss}$ for pure atomic scattering.  The lines are eqn. (\ref{widths_rs}).   {\sl Top Right}:  The r.m.s frequency shift $\Sigma$ and the FWHM $\Gamma$ are shown as a function of the $n^{th}$ surface scatter for surface motion with a Maxwellian distribution with r.m.s. speed $V^c$.  {\sl Bottom Left}:  The r.m.s frequency shift $\Sigma$ as a function of the $n^{th}$ surface scatter when both atomic thermal scattering and random cloud motion effects are combined.  The four different symbols correspond to four different values of $V^c/V^{\rm dop}$:  the triangles, circles, squares, and diamonds correspond to $V^c/V^{\rm dop}=10, \ 5, \ 3, \ 1$ respectively.  The lines are eqn. (\ref{widths_combined}).  {\sl Bottom Right}:  The FWHM $\Gamma$ as a function of $n_{ss}$;   otherwise all symbols retain the same meaning as the bottom left panel.}
\label{figure_widths}
\end{center}
\vspace{0cm}
\end{figure}
\begin{figure}
\begin{center}
\vspace{0.5cm}
\scalebox{0.5}[0.5]{
\includegraphics[angle=0]{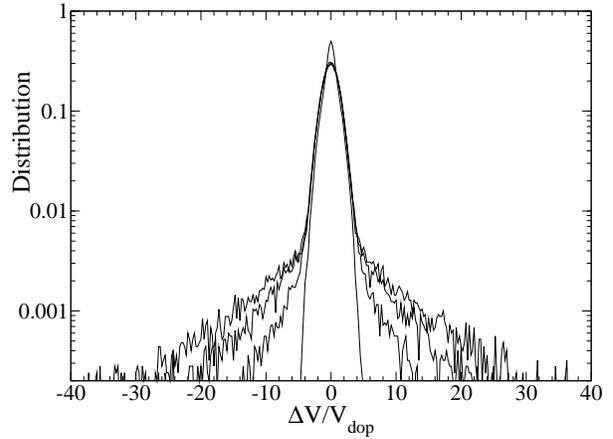} }
\caption{{\bf Repeated Surface Scatters:  Atomic Motion Only}  The normalized frequency distribution after repeated scatters off a stationary, dust-free, surface.  Shown are $n_{ss}=1, \ 5, \ 10, \ 15$, which correspond to increasing widths in the plot.  An ensemble of photons are initially injected at line center, and their frequencies are tracked after repeated cloud scatters.  For each cloud scatter, the analytic formula of \S \ref{section_surface_freqred} is used to generate the photon's exiting frequency.}
\label{figure_WidthsSpectra}
\end{center}
\vspace{0cm}
\end{figure}

\subsubsection{Resonant Scattering}
By using the analytic approximation to the surface frequency redistribution function $R(x_i,x)$ derived in \S \ref{section_surface_freqred}, we simulated repeated surface scattering for an ensemble of photons.  Carrying this out using the exact Monte-Carlo simulation is not computationally feasible.  For a few cases we compared the results of the analytic approximation with an exact Monte-Carlo simulation and a Monte-Carlo simulation that approximates the core scatterings (\S \ref{section_acceleration_scheme}), and all three are in good accord.  This gives us some confidence that the inaccuracies in the analytic approximation do not compound significantly for the number of scattering surfaces we investigated.  As shown in Figure \ref{figure_widths}, decent fits for the resonant scattering $\Sigma$ and $\Gamma$ are 
\begin{eqnarray}\label{widths_rs}
\Sigma_{\rm rs}(n_{ss})&=&\frac{1}{2} n_{ss} \ V^{\rm dop}\\
\Gamma_{\rm rs}(n_{ss})&=&3\frac{n_{ss}}{1+n_{ss}} \ V^{\rm dop} \nonumber \ .
\end{eqnarray}
The behavior can be easily understood.  If atomic thermal motions are responsible for frequency redistribution, then the line profile quickly relaxes to a Gaussian whose FWHM is determined by the width of the Doppler core.  However, the profile also acquires non-Gaussian tails from rare scattering events;  these tails dominate the r.m.s. line width $\Sigma$, hence $\Sigma\propto n_{ss}$.  For resonant scattering {\rm only}, the FWHM $\Gamma$ is a more accurate measure of the line profile than $\Sigma$.  Examples of the spectra after repeated surface scatters is shown in Figure \ref{figure_WidthsSpectra}.  

\subsubsection{Surface Motion}
As shown in \S \ref{section_kinematics}, when photons scatter off a moving surface there is a net frequency shift per surface scatter of $\langle \Delta V\rangle\sim V_{\perp}$, where $V_{\perp}$ is the velocity along the outward normal.  If the moving surfaces have a Maxwellian velocity distribution with r.m.s. speed $V^c$, then we expect that the induced r.m.s. frequency shift after $n_{ss}$ scattering surfaces will scale as $\Sigma_{\rm sm}(n_{ss})\sim \sqrt{n_{ss}} \ V^c$.  For a Gaussian distribution, the FWHM and the standard deviation are related by $\Gamma\approx 2.2\Sigma$, and so we expect $\Gamma_{\rm sm}(n_{ss})\approx 2.2\sqrt{n_{ss}} \ V^c$.  To check this, we simulated repeated surface scattering assuming an isotropic incident angle distribution and the surface scattering exiting angle distribution, eqn. (\ref{exit_angle}).  We indeed find that
\begin{eqnarray}  \label{widths_sm}
\Sigma_{\rm sm}(n_{ss})&=&\sqrt{n_{ss}} \ V^c\\
\Gamma_{\rm sm}&=&2.2\sqrt{n_{ss}}\ V^c  \nonumber \ ,
\end{eqnarray}
which is shown in the middle panel of Figure \ref{figure_widths}.  

\subsubsection{Combined Effect}
In the two bottom panels of Figure \ref{figure_widths}, we show the combined effects of resonant line broadening and surface motion.  For most multi-phase geometries, $V^c\gg V^{\rm dop}$, so line broadening is dominated by cloud motions.  In this regime, the line width can be accurately estimated by eqn. (\ref{widths_sm}).  Only if cloud motions are small and comparable to atomic thermal motions, $V^c\leqsim V^{\rm dop}$, does the behavior change.  In this case, the FWHM does not increase with the number of scatters;  instead it ``thermalizes''  to the characteristic Doppler width.  The line profile is more accurately described by eqn. (\ref{widths_rs}).  When $V^c\geqsim V^{\rm dop}$, the total r.m.s. width is (note that the dispersions add linearly rather than in quadrature)
\begin{equation}\label{widths_combined}
\Sigma(n_{ss})=2\Sigma_{\rm rs}(n_{ss}) + \Sigma_{\rm sm}(n_{ss})\ .
\end{equation}
The FWHM has a more complex behavior because the Doppler core tends to retard the increase in $\Gamma$ beyond the size of the Doppler core.

\subsection{\Lya Escape Fractions}\label{section_fesc_analytic}
In this section we show how \Lya multiphase transfer can be handled analytically based on the results in the previous subsections.  We shall first estimate the typical escape frequency of \Lya photons $\xesc$, which provides the typical surface absorption albedo $\epsilon_c(\xesc)$.   With this estimate of $\epsilon_c$ and $\N_0$, we can calculate the typical \Lya escape fraction $\fesc$ using eqn. (\ref{fesc_g=0_N0}).

We find that an adequate approximation for the typical escape frequency is given by a slight modification of the r.m.s. line width $\Sigma(n_{ss})$,  eqn. (\ref{widths_combined}), where the surface motion can be either a random Maxwellian cloud velocity with r.m.s. speed $V^c$ or a bulk outflow with typical outflow speed $V^c$ (see \S \ref{section_outflow} below).  It is {\sl not} correct to approximate the typical number of scattering surfaces $n_{ss}$ with the average number of scattering surfaces in the absence of absorption, $\N_0$,  since in general $n_{ss}\ll \N_0$ once absorption is taken into account.  A more appropriate measure of $n_{ss}$ is given by $\N$, eqn. (\ref{Nesc_N0}), the average number of scattering surfaces encountered before escape for a fixed cloud albedo $\epsilon_c$.  However, since $\epsilon_c$ is frequency-dependent, we still need to estimate the escape frequency $\xesc$.  We do so iteratively.  We first estimate $\xesc$ assuming that absorption is unimportant, 
\begin{equation}
\xesc^0=\Sigma(\N_0) \ , 
\end{equation}
which obviously overestimates $\xesc$.  We can then estimate $\N$ as:
\begin{equation}
n_{ss}\approx \N \approx (1-\epsilon_c^0) \N_0 \frac{\tanh \sqrt{2\epsilon_c^0\N_0}}{\sqrt{2\epsilon_c^0\N_0}}
\end{equation}     
where
\begin{equation}
\epsilon_c^0\equiv \frac{2 \sqrt{\epsilon(\xesc^0)}}{1+\sqrt{\epsilon(\xesc^0)}}
\end{equation}
with $\Sigma(n_{ss})$ given by eqn. (\ref{widths_combined}).  

With the above estimate of $n_{ss}=\N$, a better approximation to $\xesc$ is given by the next iteration, 
\begin{equation}
\xesc'=\Sigma(\N)\ ,
\end{equation}
which gives a better approximation for the typical cloud albedo
\begin{equation}\label{albC_lya_esc}
\epsilon_c'= \frac{2 \sqrt{\epsilon(\xesc')}}{1+\sqrt{\epsilon(\xesc')}}\ .
\end{equation}
At this point one could iterate again---using $\epsilon'_c$ to find a better approximation to $n_{ss}$--- but we find that stopping after one iteration provides escape fractions in good accord with simulations.  From  eqn. (\ref{fesc_g=0_N0}), the escape fraction given by 
\begin{equation}
\fesc\approx \frac{1}{\cosh\sqrt{2\epsilon'_c \N_0}}\ .
\end{equation}

\begin{figure}
\begin{center}
\vspace{0.5cm}
\scalebox{0.5}[0.5]{
\includegraphics[angle=0]{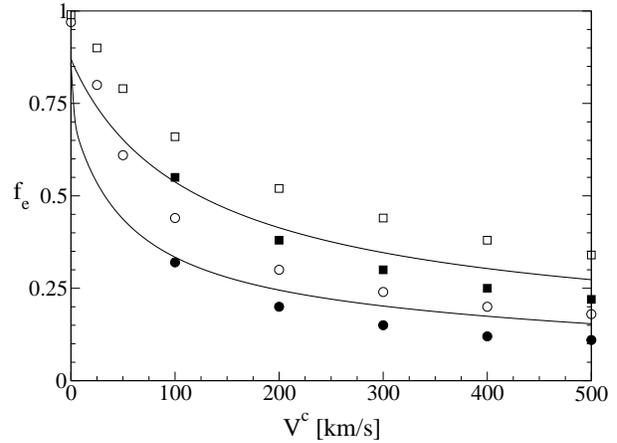} }
\caption{{\bf Analytic \Lya Escape Fraction Examples} The \Lya escape fraction $\fesc$ as a function of the typical speed of the bulk cloud motion, $V^c$.  Simulations of \Lya transfer through the random surfaces geometry (\S \ref{section_r-surfaces}) with $f_C=3$ were run for two types of bulk cloud motion: a Maxwellian velocity distribution (open symbols), and a purely radial outflow (filled symbols).  In the former case the x-axis corresponds to the r.m.s. speed of the clouds, while in the latter case the x-axis corresponds to the outflow speed, which is the same at all radii.  Two different dust contents were simulated, both with $\epsilon_d=0.5$:  $\sigma^d_{-21}=2$ (circles) and $\sigma^d_{-21}=0.2$ (squares).  In all cases the gas temperature is $10^4\Kel$.  The analytic approximations (lines) are generated by the steps outlined in text.}
\label{fig_fesc2}
\end{center}
\vspace{0cm}
\end{figure}

In Figure \ref{fig_fesc2} we compare this analytic escape fraction to simulations of radiative transfer through the random surfaces geometry (see \S \ref{section_r-surfaces}) for both random cloud motions and a bulk cloud outflow, and for several amounts of dust.  The bulk cloud outflow is purely radial, with the same speed at all radii, which approximates galactic winds outside the initial acceleration zone.  As can be seen, the analytic approximation captures the simulated escaped fraction to $\sim 20\%$ when cloud motion dominates atomic thermal motion, $V^c\geqsim 3V^{\rm dop}\sim 40\kmPs$.  As explained in \S \ref{section_line_widths}, once the effects of the Doppler core become important, the r.m.s. frequency shift $\Sigma$ is no longer a good measure of the typical frequency:  $\Sigma$ will overestimate the typical escape frequency, and hence lead to an overestimate of the absorption.  This is seen in the figure for $V^c\leqsim 40\kmPs$.   However, since the amount of absorption is typically not significant when the escape frequency is $\leqsim 40\kmPs$, for most purposes using $\Sigma$ allows one to accurately estimate $\fesc$.  Note that for bulk outflows with the same characteristic speed $V^c$, the escape fraction is smaller than for random cloud motion.  The line profile for random cloud motion is approximately Gaussian centered on the line center, with a standard deviation of $V^c$.  In contrast, the line profile for a bulk outflow is has a mean at $V^c$.  For a given $V^c$, a bulk outflow 
produces more photons far from line center than random cloud motion, and hence the bulk outflow causes more \Lya absorption.

In summary, the escape fraction depends upon five parameters:  the typical cloud speed $V^c$, the number of surface scatters in the absence of absorption $\N_0$, the gas temperature $T$, and the dust parameters $\sigma^d$ and $\epsilon^d$. 

\subsection{Dust and Gas Between the Clumps}\label{section_ICM_gas}

Thus far, we have only considered radiative transfer off opaque surfaces, and ignored absorption and/or scattering in the optically thin hot intercloud medium (ICM). We treat the hot intercloud medium as having a very low neutral HI fraction (for gas in coronal equilibrium at $T\sim 10^{6}$K, $x_{HI} \sim 10^{-5}$), as well as being dust-depleted, due to sputtering and other dust-destruction processes. ICM resonant scattering and absorption is easily incorporated in our Monte-Carlo simulations at little computational cost, and we have experimented with various prescriptions for the ICM in our runs. In most cases, we find that radiative transfer within the ICM can be neglected, and here we show some simple estimates demonstrating why this is the case.

Let us first consider dust absorption in the ICM, assuming that resonant scattering in the ICM is negligible. It is easy to show that for a multiphase medium, the relative fractional column densities of a species $i$ in phases $X,Y$ is simply given by the relative mass fraction of species $i$ in that phase:
\begin{equation}
\frac{N_{X}^{i}}{N_{\rm tot}^{i}}= \frac{f_{M,X}^{i}}{f_{M,Y}^{i}} = \left( \frac{f_{M,X}}{f_{M,Y}} \right) \left( \frac{x_{X}^{i}}{x_{\rm tot}^{i}} \right)
\end{equation}
where $x^{i}_{X}$ is the mass fraction of species $i$ in phase $X$. Thus, for instance, $N_{\rm HI}^{c}/N_{\rm HI}^{\rm ICM} \approx f_{M,c} [x_{\rm HI}^{\rm ICM}]^{-1} \gg 1$ for $x_{\rm HI}^{\rm ICM} \ll 1$, and the observed HI column density $N_{\rm HI,21}^{\rm obs} \sim 1$ is strongly dominated by gas in the cold phase.

We can use this to estimate scattering in the ICM. Due to its random walk as it scatters off optically thick surfaces, a Ly$\alpha$ photon traverses an ICM column density $\sim \sqrt{\cal{N}}$ times larger than for a straight line path. At a characteristic escape frequency $V_{2}^{e} \sim 1$, it therefore encounters an ICM HI optical depth:
\begin{equation}
\tau_{\rm HI}^{\rm ICM}=0.07 N_{\rm HI,21}^{\rm tot} \left( \frac{f_{\rm M,ICM}}{0.7}\right) \left( \frac{\cal{N}}{5} \right)^{1/2} \left( \frac{x_{\rm HI}^{\rm ICM}}{10^{-4}} \right) [V_{2}^{e}]^{-2}.
\end{equation} 
Hence, resonant scattering in the ICM is negligible. It only becomes important when the photon is within the Doppler core $V_{2} \sim 0.26$ for the parameters above). However, even in this rare case $\tau_{\rm HI}^{\rm ICM}\sim$few, comparable to the number of surface scatterings $\cal{N}$.

What about dust absorption in the ICM? Let us suppose that due to dust depletion, only $f_{d}^{\rm ICM} \sim 0.05$ of all the dust in the galaxy is in the ICM. If the total dust optical depth through the cold phase is $\tau_{d}^{c} \sim 1 \, N_{\rm HI, 21}^{\rm tot} \sigma_{-21}^{a}$, then the total dust optical depth through the ICM is $\tau_{d}^{\rm ICM} \sim f_{d}^{\rm ICM} \tau_{d}^{c}$. The mean free path of a Ly$\alpha$ photon is $l \sim r_{\rm gal}/\sqrt{\cal{N}}$. Hence, between each bounce, a Ly$\alpha$ photon traverses an optical depth $\tau_{a}^{\rm bounce} \sim \tau_{d}^{\rm ICM}/\sqrt{\cal{N}}$, and has a probability of absorption in the ICM of $P_{\rm absorb}^{\rm ICM} \approx 1- e^{-\tau_{a}^{\rm bounce}} \approx \tau_{a}^{\rm bounce}$, or:
\begin{equation}
P_{\rm absorb}^{\rm ICM} \approx 0.02 \left( \frac{\cal{N}}{5} \right)^{-1/2} \left( \frac{f_{d}^{\rm ICM}}{0.05} \right) \left( \frac{\tau_{d}^{c}}{1.} \right)
\end{equation}
By contrast, during each surface scatter, the photon has an absorption probability of:
\begin{equation}
P_{\rm absorb}^{c} = \epsilon_{c} \approx 2 \sqrt{\epsilon(x_{e})} \approx 0.12 V_{2}^{e} [\sigma_{-21}^{a}]^{1/2}.
\end{equation}
Thus, photons are more likely to be absorbed on cloud surfaces, rather than in the ICM, justifying our neglect of ICM dust absorption.  

Obviously, all of these statements are parameter dependent and there {\it are} cases when scattering and absorption in the ICM cannot be neglected (if the IGM dust or HI content is high). For instance, if the HI fraction in the ICM is high, then \Lya can be strongly quenched. This may be partly responsible for the variation in the \Lya equivalent widths amongst different galaxies (see discussion in \S\ref{section:EW}). 

\subsection{Accelerated Radiative Transfer on a Grid}\label{section_accelerated_mc}

The approach of this paper is to identify all optically thick surfaces in a Monte-Carlo simulation and apply the scattering properties identified in \S\ref{section_surfaces} to them, thus affording vast computational speed-ups. This should be readily amenable to radiative transfer in numerical simulations. For this scheme to be accurate: (i) the surfaces must be sufficiently optically thick that transmission is negligible; i.e., they should satisfy equation (\ref{Npt_lya}). (ii) The approximation that the photon is either reflected or absorbed on the spot without significant spatial diffusion must hold. We now discuss this second requirement. 

For the "on the spot" approximation to hold, the photon's mean free path $\ell_{i}$ should be significantly smaller than the grid cell size $L_{\rm cell}$. The photon typically moves a distance $\sim \sqrt{\cal{N}^{\rm surf}} \ell_{i} \sim 5 \ell_{i}$ (see \S\ref{section_ScatNumLya} for discussion of $\cal{N}^{\rm surf}$) whilst scattering within the surface. Thus, we require $\ell_i\leqsim \alpha\ L_{\rm cell} $ where $\alpha\sim 1/10$ is a constant that designates the desired level of accuracy for the approximation. The "on the spot approximation" is accurate if the HI density is larger than a critical density:
\begin{equation}
n_{\rm cell}^{\rm HI}\geq n^*= \frac{13}{(\alpha/0.1)} \frac{{V^{i}_{2}}^2}{(L_{\rm cell}/\pc)}\ \cm^{-3}
\end{equation}
where $V^i$ is the incident \Lya frequency shift off line center in the rest frame of the HI in the cell (in velocity units). 

Alternatively, if one is willing to sacrifice some spatial resolution, one can consider a group of neighboring cells which are opaque across their total length (and thus satisfy equation (\ref{Npt_lya})) even though an individual cell may not be opaque.  For cubic blocks of $N_{\rm blk}$ cells per side, the entire block surface will act like an absorbing mirror if
\begin{equation}
\langle n\rangle_{\rm blk}^{\rm HI} \geq n^*= \frac{13}{(\alpha/0.1)} \frac{{V^{i}_{2}}^2}{(N_{\rm blk} L_{\rm cell}/\pc)}\ \cm^{-3}
\end{equation}
where $\langle n \rangle_{\rm blk}$ is the average HI density within the block.  The surface approximations break down if the block is strongly inhomogeneous, i.e., the mean free path changes over a length scale that is much shorter than the block length. This is equivalent to  $| n_{\rm cell}/\nabla(n_{\rm cell}) |\leq \beta N_{\rm blk} L_{\rm cell}$, where $\beta\sim 1/10$ is a constant that designates the desired level of accuracy for the approximation.  In terms of cells on a grid, let $n(2)_{\rm cell}$ and $n(1)_{\rm cell}$ be any two neighboring cells in the block.  The second condition on the absorbing mirror approximation takes the form 
\begin{equation}
\left| \frac{n(2)_{\rm cell}}{n(1)_{\rm cell}}-1\right| \leq  0.1\frac{(\beta/0.1)}{N_{\rm blk}}\ .
\end{equation}

We look forward to implementing these ideas in numerical simulations in the future.

\section{Applications}
\label{section:applications}

We briefly discuss two examples of applications of our radiative transfer framework. Many more extensive studies are possible. 

\subsection{\Lya Equivalent Widths}
\label{section:EW}
Most \Lya photons are produced in the HII regions surrounding sources of ionizing radiation, where roughly $2/3$ of the ionizing photons are converted into \Lya photons (under case B recombination). Hence, in the absence of radiative transfer effects, the equivalent width measures the number of ionizing photons emitted relative to the UV continuum near $1200\Ang$. There are numerous examples of high-redshift sources which have equivalent widths which are too large to be produced by conventional stellar populations. About $\sim 2/3$ of the SCUBA submm galaxies with accurate positions from radio detections have \Lya in emission, many with equivalent widths too great for stellar sources \citep{smailetal}. The mysterious \Lya emitters at $z\sim 3.1$ observed by \citet{steidel} have enormous \Lya fluxes, but no observed continuum. Finally, the LALA survey detects many high redshift $z=4.5,5.7$ sources with equivalent widths ${\rm EW} \ge 150 \, {\rm \AA}$ significantly in excess of any known nearby stellar population \citep{rho03}. An AGN origin is unlikely because follow-up observations show no signs of the X-rays and high-ionization lines expected for a type II quasar source \citep{Wang04, Dawson04}. Another possibility is that the \Lya emission is due to an extremely top-heavy population of massive PopIII stars. However, there are no signs of the strong HeII emission at $1640\Ang$ expected from metal-free stars \citep{Dawson04}.  

Another possibility for high \Lya EWs, originally suggested by \citet{neufeld91}, is radiative transfer effects. If the continuum is more absorbed than \Lya photons during the escape from the host galaxy, then the equivalent width of the transmitted spectra is {\sl larger} than the equivalent width of the source. The initial and transmitted equivalent widths are basically related by the ratio of \Lya to continuum escape fractions, 
\begin{equation}
\EW_{\rm out}\sim \frac{\fesc^{{\rm Ly}\alpha}}{\fesc^{\rm ctm}} \ \EW_{\rm src}\ ,
\end{equation}
where $\EW_{\rm src}$ is the source equivalent width and $\EW_{\rm out}$ is the equivalent width for the escaping photons. In order for a ``normal'' starburst IMF with an intrinsic equivalent width of $\EW_{\rm src}\sim 150\Ang$ to produce an observed equivalent width of $\EW_{\rm out}\geqsim 300\Ang$, then radiative transfer must account for a ``boost''  by a factor of at least 2---3. For sources where no continuum is observed, the continuum must be preferentially extinguished by an even larger factor.

\begin{figure}
\psfig{file=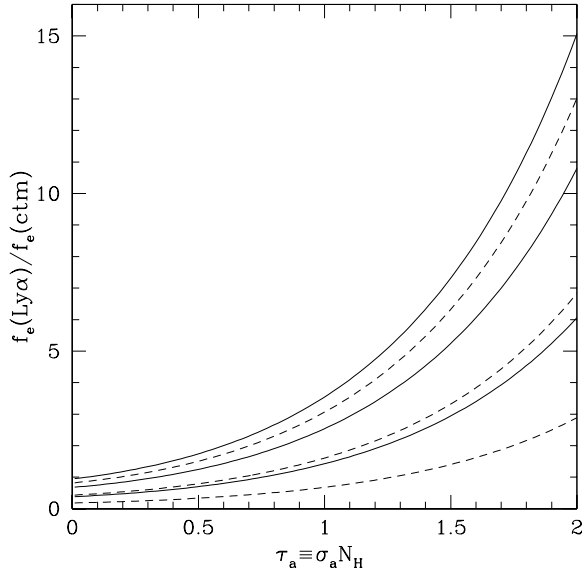,width=80mm}
\caption{{\bf \Lya/Continuum Escape Fraction Ratio}  The \Lya to continuum escape fraction ratio, $\fesc^{{\rm Ly}\alpha}/\fesc^{\rm ctm}$, as a function of the total dust absorption optical depth $\tau^a_{-21}=\epsilon_d \sigma^d_{-21} N_{21}$, assuming the temperature of the neutral phase is $\sim 10^{4}$K. From top to bottom, ${\cal{N}}_{0}$ is $(1,4,10)$, for bulk gas motions of $50 \, {\rm km \, s^{-1}}$ (solid lines), and $250 \,{\rm km\,s^{-1}}$ (dashed lines) respectively. For these parameters, the \Lya escape fractions are $\fesc^{{\rm Ly}\alpha}=(0.94,0.68,0.38)$, and $\fesc^{{\rm Ly}\alpha}=(0.81,0.43,0.18)$ respectively, independent of the total dust optical depth $\tau_{a}$. Most of the EW boost comes from the low escape fractions for continuum photons under optically thick conditions; very approximately, $\fesc^{\rm ctm} \sim {\rm exp}(-\tau_{a})$.}
\label{figure_EW1}
\end{figure}

Let us now estimate equivalent width boosts in our multi-phase model to see if this is possible. For any multiphase medium where the gas resides in clumps that are very opaque to \Lya, the surface scattering approximations apply, and so the \Lya escape fraction can be analytically estimated as in \S \ref{section_fesc_analytic}.  What about the continuum escape fraction? For simplicity, assume that each gas clump is not opaque to dust extinction:  $\tau^d_c\leqsim 1$, where $\tau^d_c$ is the dust extinction (scattering+absorption) optical depth across a clump diameter\footnote{However, the total dust absorption optical depth across the entire galaxy (many clumps) can be significantly greater than unity.}.  Since the self-shielding effect of clumpy gas is therefore small for the continuum photons, the effective dust distribution is approximately homogenous for the continuum radiative transfer.  The escape fraction for a photons injected in the middle of a homogenous medium, with an absorption albedo $\epsilon^d\approx 1/2$, is approximately that given by eqn. (\ref{fesc_g=0_N0}), 
\begin{equation}\label{fesc_ctm_thin}
\fesc^{\rm ctm}\approx 1/\cosh\left[\sqrt{\epsilon^d\left((\tau^d)^2+2\tau^d\right)}\right]\ ,
\end{equation}
where $\tau^d\equiv N\ \sigma^d$ is the average dust extinction optical depth through a region with average HI column density $N$.  Figure \ref{figure_EW1} shows how the ratio $\fesc^{{\rm Ly}\alpha}/\fesc^{\rm ctm}$ varies as a function of $\N_0$ for a fiducial set of multiphase gas parameters. A substantial ``boost'' in the transmitted equivalent width due to selective absorption of the continuum is quite reasonable, as long as two basic conditions are met:  1) there must be enough dust present to absorb a substantial fraction of the continuum.  2) this dust must be pre-dominantly located in dense neutral gas, so that the \Lya photons are shielded from absorption.  The latter condition is discussed in \S \ref{section_ICM_gas} above, and so we turn to discussing the first condition.

The \Lya escape fraction depends only weakly on the overall dust content of the galaxy. In Fig \ref{figure_EW1}, $\fesc^{\rm Ly\alpha} \sim 0.2-0.9$ over a wide range of parameters, with the escape fraction decreasing with the number of surface scatters ${\cal N}_{0}$ and bulk gas motion $V_{c}$. On the other hand, because continuum photons are not shielded from dust by resonant scattering, they see the full optical depth of dust absorption, and very approximately, $\fesc^{\rm ctm} \approx {\rm exp}(-\tau_{a})$. A significant boost in the equivalent width therefore requires that $\tau^d \ge 1$.  If the average HI column density across the region is $\langle N\rangle$, then $\tau^d\sim1$ requires $\sigma^d_{-21}\sim 1/\langle N_{21}\rangle$.  A damped \Lya type system with an average column density $\langle N\rangle\sim 10^{22}\cm^{-2}$ would require a dust extinction cross-section per hydrogen of $\sigma^d\sim 10^{-22}\cm^2/H$, which roughly corresponds to a metalicity of $\sim 1/10$ solar. For sources in which these differential radiative transfer effects are taking place, the equivalent width should statistically have a positive correlation with the FIR flux.  This correlation could potentially break down in more developed galaxies at lower redshifts, where the \Lya shielding effect of the HI can be broken by higher gas speeds in deeper potential wells (rendering clumps optically thinner in \Lya), as well as the build-up/survival of dust in low density inter-clump gas.  

\subsection{Multi-phase Outflows}\label{section_outflow}
We turn to discussing the effect of a multi-phase gas outflow (or inflow) on the \Lya emission line profile.  The effects of an outflowing shell (e.g. \cite{Tenorio-Tagle99, ahn03, ahn04}) and a Hubble like expansion of a uniform gas sphere (e.g. \cite{loeb99, zheng2}) on the \Lya emission line is a well studied problem.   In both the expanding shell and expanding sphere scenarios, the generic effect is that a characteristic outflow speed $V^f$ produces a redshifted emission peak at $V^{\rm peak}\sim -V^f$, with an asymmetric shape that has a longer tail on the red side of the peak.  The peak comprises photons that reflect off the far side of the expanding gas, which Doppler shifts the frequency by $\sim -V^f$ (see \S \ref{section_kinematics}).  However, in order for these singly back-scattered photons to escape, the intervening gas column density must be small enough for the photons to be transmitted, rather than be reflected a second time.  In the rest frame of the near-side shell, the singly back-scattered photons have a frequency shift ${V^{\prime}}^{\rm peak}\sim -2V^f$.  In \S \ref{section_Npt_lya}, we showed that a non-negligible amount of \Lya photons will be transmitted through a slab if $N_{21}<N^{\rm pt}_{21}\approx 0.05 [V_2]^2 [\sigma_{-21}^{a}]^{-1}$. Setting $V\sim -2V^f$,  we see that an outflow with speed of $200\kmPs$ will only allow a non-negligible amount of singly back-scattered photons to be transmitted if the intervening column density is $N\leq 2\times10^{20}\cm^{-2}$.  Observational estimates of the column density in galactic winds often exceed this, yet \Lya is still often seen.  

The main distinguishing feature of a multiphase outflow is that it allows photons of any frequency to escape even when the intervening gas column depth is very large, $N\geq N^{\rm pt}$.  As in the homogeneous gas outflow models with smaller column densities $N<N^{\rm pt}$, we find that for multiphase outflows, the emission peak is redshifted by $\sim V^f$.  However, emission is still detectible even when $N\gg N^{\rm pt}$, as expected.    

In particular, we investigated the emission profile for two basic types of multiphase outflow geometries: an outflowing shell with holes (\S \ref{section_shell_holes}) and an outflowing ensemble of gas clumps modeled with the Random Surfaces geometry (\S \ref{section_r-surfaces}).   In both cases, all surfaces were given a radially velocity with constant speed.  This choice is meant to reflect galactic winds, where the gas reaches the asymptotic wind speed quickly.  We placed a source of line center photons in the center of the region. Since the regime of optically thick gas has been given the least attention, we assume that the extreme case holds, where none of the photons penetrate through the gas.  In this limit the surface scattering approximations of \S \ref{section_SingleClouds} apply in the rest frame of the scattering surface.  The kinematics of Doppler shifts in and out of a moving surface can be found in \S \ref{section_kinematics}.  In order to distinguish the effects of outflow from the effects of random bulk gas motion, we assume that there is no random bulk motion, so that each gas surface has an exactly radial velocity, $\vec{V}^s=V^f \hat{r}$.

\subsubsection{Outflowing Shell with Holes}
\begin{figure}
\begin{center}
\vspace{0.5cm}
\scalebox{0.5}[0.5]{
\includegraphics[angle=0]{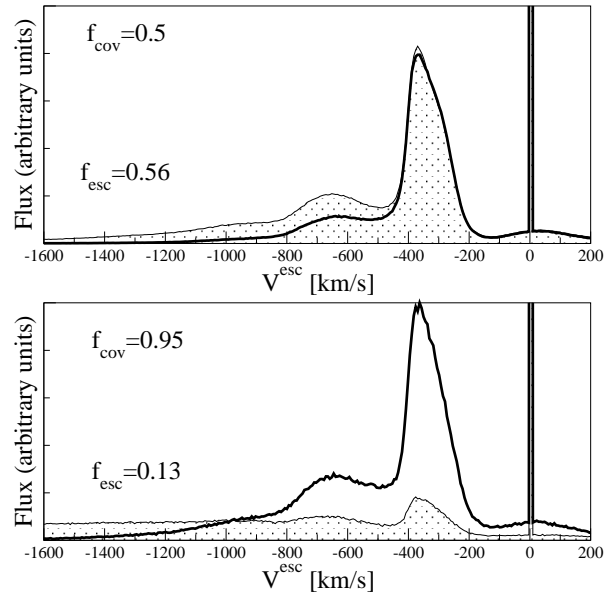} }
\caption{{\bf Outflowing Shell with Gaps}  The normalized emission profile as a function of the frequency shift off line center, for an outflowing shell with speed $V^f=200 \kmPs$.  Two dust amounts were simulated: $\sigma^a=0$ (thin line, filled in) and $\sigma^a_{-21}=1$ (thick line).  In all simulations, we used a gas temperature $T_4=1$.  For high redshift galaxies, the blue side of the profile would be quenched by IGM absorption.  The spike of photons that escape at line center is easily scattered out of the line of sight by a small ($N\geq 5\times 10^{13}\cm^{-2}$) intervening column density of HI, and thus not likely to be observed.}
\label{figure_shell_outflow}
\end{center}
\vspace{0cm}
\end{figure}
For photons perpendicularly incident on the inner side of the shell, the net frequency shift induced by a moving scattering surface is $\Delta V=-\frac{5}{3}V^f$ (see \S \ref{section_KinematicsVelocity}).  Since most photons escape after encountering either zero or one scattering surface, the two dominant peaks in the emission profile will be at $V_{(0)}=0$ and $V_{(1)}\approx -\frac{5}{3}V^f$.  The maximum frequency shift after encountering one scattering surface is $\Delta V_{\rm max}=-2V^f$, and so the emission peak at $V_{(1)}$ should have a sharp cut-off at $V=-2V^f$.  These emission peak features are verified in Figure \ref{figure_shell_outflow}.  The series of secondary peaks are at integer multiples of $V_{(1)}$, but tend to become blended together to form an long red-side tail to the profile. A possible exception is the third peak at $V_{(2)}=2V_{(1)}=-\frac{10}{3}V^f$ (composed mostly of photons that encounter exactly two scattering surfaces before escaping), which can also be prominent in the profile.

\subsubsection{Outflowing Clumps}
\begin{figure}
\begin{center}
\vspace{0.5cm}
\scalebox{0.5}[0.5]{
\includegraphics[angle=0]{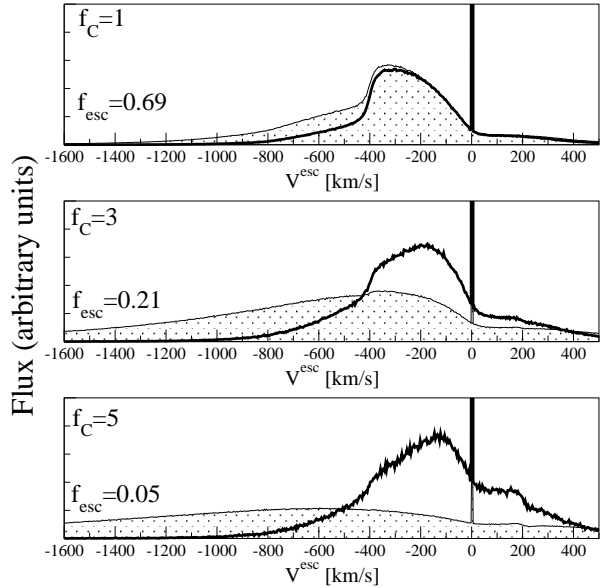} }
\caption{{\bf Outflowing Clumps} The normalized emission profile as a function of the velocity shift from line center.  The thick lines are $\sigma^a_{-21}=1$ while the thin lines (filled in) are dust free, $\sigma^a=0$.  The gas temperature and outflow speed are the same as in Figure \ref{figure_shell_outflow}.  From top to bottom, each panel is a different covering factor: $f_C=1, \ 3, \ 5$.  The escape fractions for the $\sigma^a_{-21}=1$ simulations are indicated.  The delta function emission spike at $V^{\rm e}=0$ is composed of all the photons that escape freely without striking a clump.  As noted in Figure \ref{figure_shell_outflow}, exact line center photons are likely to scattered out of the line of sight before being observed.}
\label{figure_outflow_rs}
\end{center}
\vspace{0cm}
\end{figure}

\begin{figure}
\begin{center}
\vspace{0.5cm}
\scalebox{0.5}[0.5]{
\includegraphics[angle=0]{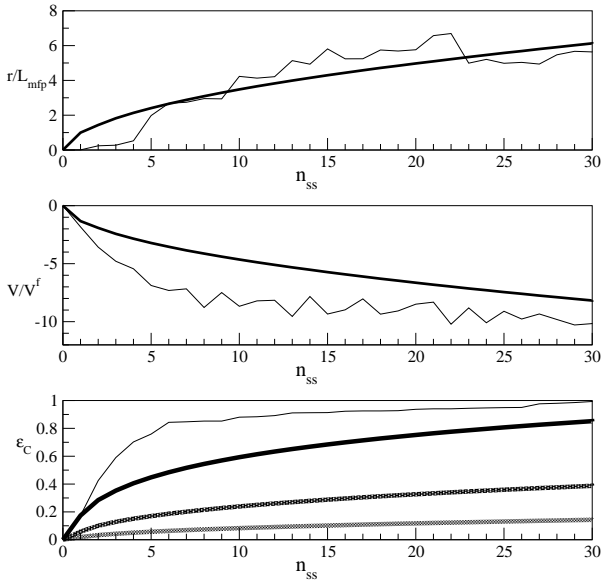} }
\caption{{\bf Random Surfaces Bounce History}  {\sl Top}:  The radial position $r$ (in units of the mean free path between surfaces $L_{\rm mfp}$), plotted against the number of surface scatters $n_{ss}$.  The thin line is a representative single photon history, the thick solid line is the average over an ensemble of photons.  {\sl Middle}:  The frequency shift $V$ (in units of the outflow speed $V_{f}$), plotted against $n_{ss}$.  The thin solid line is a representative photon history, the thick solid line is the average history.  {\sl Bottom}:  The \Lya surface absorption albedo $\epsilon_c$ is plotted against $n_{ss}$, for $V^{\rm f}=200\kmPs$ and a gas temperature $T_4=1$.  We use the approximation $\epsilon\approx \beta/\Phi(x)$, which breaks down when $\epsilon\approx 1$.  The thin solid line is a representative photon history for $\beta=10^{-8}$.  The thick solid lines are the average histories for several different values of the dust content $\beta$.  From lightest to darkest, $\beta=10^{-10}, \, 10^{-9}, \, 10^{-8}$. }
\label{figure_history}
\end{center}
\vspace{0cm}
\end{figure}

For a simple model of outflowing gas clumps we used the Random Surfaces geometry, which is described in detail in \S \ref{section_r-surfaces}.   In Figure \ref{figure_outflow_rs}, we show how the profile varies with the covering factor, for dust free gas $\sigma^a=0$ and for a Milky Way type dust content $\sigma^a_{-21}=1$. As for the case of an outflowing shell with holes, the inclusion of dust suppresses photons which have redshifted far from line center (so the HI no longer shields them from the dust), and sharpens the line profile. To provide some insight into how a clumpy outflow causes a redshift in the emission line, in Figure \ref{figure_history} we show how the photon radius, frequency, and cloud absorption albedo vary as a function of $n_{ss}$.  

\section{Conclusions}

Our main technical, radiative transfer results are:
\begin{itemize}
\item  With the aid of Monte-Carlo simulations, we study the scattering properties of \Lya photons incident on an opaque, dusty, moving cloud. We derive fitting formulas for the absorption probability, frequency and angular redistribution functions of incident photons.
\item  These formulas can be incorporated into radiative transfer codes, affording a vast computational speed up, and making feasible otherwise intractable calculations.
\item  Analytically, a multiphase gas geometry can be accurately characterized by a single number, $\N_0$, the number of surface scatters in the absence of absorption. Other factors---such as the cloud radii distribution for fixed $\N_0$---are generally unimportant.
\item  We derive analytic formulas for the \Lya escape fraction and line widths.
\item  Several archetypal geometries are explored: randomly placed spherical clouds, randomly placed surfaces (an abstraction of the prior geometry),  a shell with holes, and an open-ended, cylindrical cavity.
\item  Constant speed, radial, outflows are analyzed for broken shells and random surfaces.  The red-shifted peaks and widths are connected to the geometry and outflow speed. 
\end{itemize}
  
Our main results of direct observational relevance are:
\begin{itemize}
\item  \Lya can escape from multi-phase dusty galaxies for HI column densities where it would be strongly quenched in a single-phase medium. 
\item If most of the dust resides in a neutral phase which is optically thick to Ly$\alpha$, the \Lya equivalent width can be strongly enhanced: while \Lya photons typically scatter off such surfaces (which shield the dust), continuum photons penetrate inwards and are preferentially absorbed. 
\item  When the characteristic bulk gas speed exceeds $\sim 100\kmPs$, the \Lya line width is dominated by the gas motion, and resonant scattering frequency redistribution is sub-dominant effect.
\item  Multiphase outflows generically produce \Lya line profiles that have the characteristic asymmetric shape seen in many starburst galaxies and \Lya emitters.
\item  Multiphase outflows can produce line widths several times larger than the actual outflow speed.
\end{itemize}

The ISM of galaxies at both low and high redshift is almost certain to be both dusty and multi-phase: metal and dust production begin very early, given the short lifetime of massive stars, and thermal instability is almost inevitable under galactic conditions. Nonetheless, despite an extensive literature, to the best of our knowledge this is the first detailed numerical study of resonance-line radiative transfer in a multi-phase dusty medium. The ground is surprisingly rich, and many future applications are envisaged!

\section*{acknowledgments}
We thank Lars Bildsten, Andrew Blain, Joss Bland-Hawthorne, Michael Fall, Sangeeta Malhotra, Crystal Martin, David Neufeld, James Rhoads, Mike Santos, Alice Shapley for helpful conversations. This work was supported by NSF grant AST0407084.

\appendix
\section{Tests of the Code}\label{section_tests}
In this Appendix we show various tests of our Monte Carlo code.  First, we test the code against known analytic solutions for optically thick slabs.  Second, we compare the acceleration scheme described in \S \ref{section_acceleration_scheme} to exact simulations. 

\subsection{Comparison to Analytic Solutions}
We tested the exact Monte Carlo code against known analytic solutions for very opaque slabs with a source at the mid-plane.  First, we compared the emission frequency profile when the slab is pure HI, so there is no absorption.  Second, we compared the escape fractions when the slab contains a small amount of absorbing dust, where the absorption cross-section of the dust is frequency independent. 

For an optically thick ($a\tau_{h0} > 10^3$), uniform slab without dust, where $\tau_{h0}$ is the center-to-surface hydrogen optical depth at line center\footnote{In many papers on this subject, e.g. \cite{harrington73} and \cite{neufeld90}, ``$\tau_0$'' refers to the mean optical depth, while we use the line center optical depth.  In our notation, $\phi(x)\tau_0=\Phi(x)\tau_{h0}$ where $\phi(x)$ is the {\sl normalized} Voigt profile, which means $\tau_0=\sqrt{\pi}\tau_{h0}$ when $a\ll1$.}, \cite{harrington73} derived the mean intensity emission spectrum, for line center photons which are injected at the slab center (see \cite{neufeld90} for various extensions)
\begin{equation} \label{N90:J}
J(\pm\tau_{h0}, x)=\frac{6^{1/2}}{24\pi^{1/2}}\frac{x^2}{a\tau_{h0}} \left\{\cosh\left[\frac{\pi^{3/2}}{54^{1/2}}  \frac{|x|^3}{a\tau_{h0}}\right]\right\}^{-1}\ ,
\end{equation}
\begin{figure}
\scalebox{0.5}[0.5]{
\includegraphics[angle=0]{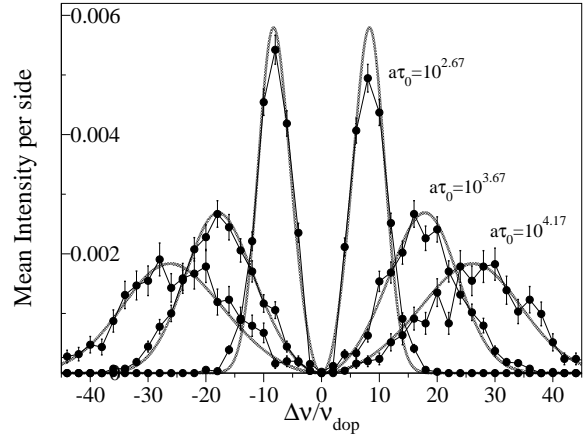}  }
\caption{\label{slab_spectra}\textbf{Spectra Emergent from Slab, Central Source.}  The surface mean intensity $J(x)$ as a function of frequency, for a central source of \Lya photons in an optically thick slab.  For three different values of $a\tau_{h0}$ (labeled in the plot), the analytic surface intensity (grey line) is compared to exact Monte-Carlo simulations (circles).  The optical depth is fixed at $\tau_{h0}=10^6$.   Three different gas temperatures are used: $T=10\Kel, \ 10^2\Kel, \ 10^4\Kel$, which corresponds to $a\tau_{h0}=10^{4.17}, \ 10^{3.67}, \ 10^{2.67}$, respectively.  The relative error for each frequency bin is approximately $1/\sqrt{N_{bin}}$, where $N_{bin}$ is the number of photons in the bin.} 
\end{figure}
where $J$ is the mean intensity and $\tau_0$ is the line center hydrogen optical depth from center-to-surface.  Figure \ref{slab_spectra} compares the simulation to the formula for slabs at three different temperatures. 

For an optically thick slab ($a\tau_{h0}>10^3$) with a center-to-surface absorption optical depth $\tau_a$, \cite{neufeld90} derived the exact escape fraction for a central source, in the limit $(a\tau_{h0})^{1/3}\gg\tau_a$, as well as several approximation formulae.  In particular, the escape fraction is well approximated by 
\begin{equation}\label{N90:ef}
\fesc\approx 1/\cosh\left[ \frac{3^{1/2}}{\pi^{5/12}\zeta}\, \left(a\tau_{h0}\right)^{1/3} \tau_a \right]\ ,
\end{equation}
where $\zeta$ is an order unity fitting parameter.  \cite{neufeld90} found that the choice $\zeta=0.525$ gives a good approximation to the exact analytic escape fraction.  
\begin{figure}
\scalebox{0.5}[0.5]{
\includegraphics[angle=0]{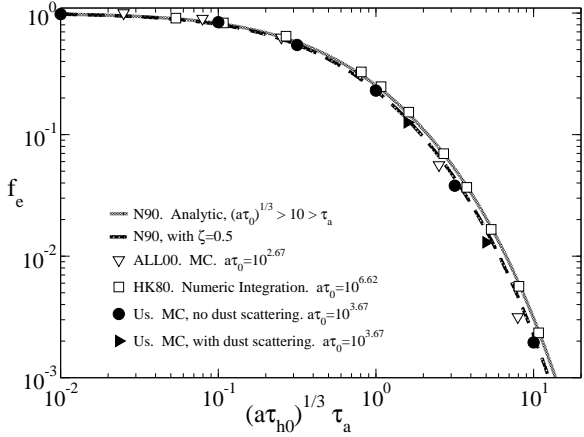}  }
\caption{\label{slab_ef}\textbf{Dusty Slab Escape Fraction, Central Source.}  The escape fraction from the middle of an optically thick, dusty slab is compared for several analytic and numerical methods.  As has been shown analytically, the escape fraction depends mainly upon the combination $(a\tau_{h0})^{1/3}\tau_a$.  Including dust scattering, using $\epsilon_d=0.5$ and $g_d=0.6$, did not significantly affect the escape fraction.  
} 
\end{figure}
For comparison, we have included in Figure \ref{slab_ef} the escape fractions found by \cite{hummer80} using numerical integration techniques and the escape fractions found by \cite{ahn00} using a Monte Carlo simulation similar to ours.   We find that the choice of fitting parameter $\zeta=0.5$ gives a good approximation to our Monte Carlo results.  Both \citet{ahn00} and our Monte Carlo simulations show slightly more absorption than the analytic formula from \citet{neufeld90}.  The analytic treatments assume the Lorentzian wing profile all the way down to $x=0$, neglecting the Gaussian core.  This will underestimate the number of scatters spent in the core.  Although the absorption probability per interaction is small in the core,  neglecting core bounces will cause a slight underestimate of the overall absorption probability. 

\subsection{Testing the Acceleration Scheme}\label{section_tests_exact}
\begin{figure}
\begin{center}
\vspace{0.5cm}
\scalebox{0.5}[0.5]{
\includegraphics[angle=0]{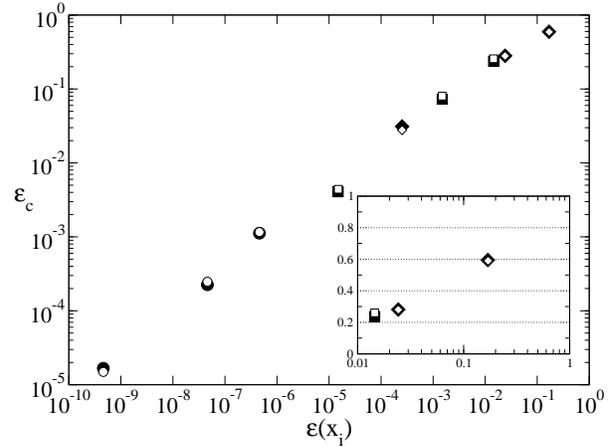} }
\caption{{\bf Accelerated Monte Carlo Comparison} The net surface absorption probability $\epsilon_c$ is shown as a function of the incident absorption probability per scatter $\epsilon$, for \Lya photons incident on a semi-infinite plane of dusty HI at $10^4\Kel$.  The dark symbols are simulations that use the accelerated Monte Carlo scheme while the white symbols are exact simulations.  The circles, squares, and diamonds are for incident frequencies $x_i=1, \, 5, \, 20$, respectively.  The dust has an absorption albedo $\epsilon_d=0.5$ and a scattering asymmetry parameter $g_d=0.5$.  For each incident frequency, simulations for three values of $\sigma^a$ were run.  For each frequency, from left to right (or equivalently, bottom to top), the dust values are $\sigma^a_{-21}=0.01,\,1, \, 100$.}
\label{figure_ec_comp}
\end{center}
\vspace{0cm}
\end{figure}

We tested the acceleration scheme, described in \S \ref{section_acceleration_scheme}, by comparing the surface absorption probability against exact simulations.  Exact simulations are computationally expensive, and so we could only test the acceleration scheme against a handful of exact cases.  As shown in Figure \ref{figure_ec_comp}, the acceleration is quite accurate, even for initial frequencies in the line core.     

\section{Surface \Lya Frequency Redistribution Formula}\label{appendix_Redistribution}
In this appendix we first show that $R(\tilde{x}_i, x;\alpha)$ given in eqn. (\ref{R_fit}) has a unit norm over $x$, as claimed.  Second, we outline the steps used to derive eqn. (\ref{exiting_x}), the generating function for $R(\tilde{x}_i, x; \alpha)$.   

To integrate $R(\tilde{x}, x;\alpha)$ over $x\in(-\infty, \infty)$, first change variables to $u\equiv x^3-\tilde{x}_i^3$.  Then the integral becomes
\begin{equation}\label{R_int_1}
\int_{-\infty}^{\infty}{\rm d}x \ R(\tilde{x}_i, x; \alpha)=\frac{\tilde{x}_i^2\sqrt{\alpha}}{\pi} \int_{-\infty}^{\infty}{\rm d}u \ \frac{1}{\alpha \tilde{x}_i^4 + u^2} 
\end{equation}
The integral over the functional form $(A+u^2)^{-1}$ is a standard integral:
\begin{equation}\label{Cauchy_int}
\int {\rm d}u \ \frac{1}{A+u^2}= \frac{1}{\sqrt{A}}\ \arctan\left(\frac{u}{\sqrt{A}}\right)\ .
\end{equation}
Use of this formula in eqn. (\ref{R_int_1}) shows that the integral over $x$ equals one, and so $R(\tilde{x}_i, x; \alpha)$ is correctly normalized over $x$ as advertised.  

To randomly generate exiting frequencies $x$ that obey the probability distribution $R(\tilde{x}_i, x;\alpha)$, we use the transformation method \citep{Press92}.
First, select a random univariate $u\in[0,1]$ and set
\begin{equation}
u=\int_{-\infty}^x{\rm d}x' \ R(\tilde{x}_i, x';\alpha)\equiv F(x)\ .
\end{equation}
The frequency $x$ is then given by functional inversion $x=F^{-1}(u)$.  As above, the integral is best carried out by changing variables to $u\equiv x^3-\tilde{x}_i^3$.  Then $F(x)$ is given by eqn. (\ref{R_int_1}), except that the upper limit is ``$x^3-\tilde{x}_i^3$'' rather than ``$\infty$''.  By using equation (\ref{Cauchy_int}) to complete the integration, we find that 
\begin{equation}
F(x)=\frac{1}{\pi}\left( \arctan\left(\frac{x^3-\tilde{x}_i^3}{\tilde{x}_i^2\sqrt{\alpha}}\right) + \frac{\pi}{2}\right)\ .
\end{equation}
The functional inversion gives the randomly drawn exiting frequency $x$:
\begin{equation}
x=\left(\tilde{x}_i^3-\tilde{x}_i^2 \tan\left(\pi u\right)\right)^{1/3}\ .
\end{equation}

\section{1-D Transfer for an Arbitrary Scattering Asymmetry Parameter}\label{section_1Dg}
\begin{figure}
\begin{center}
\vspace{0.5cm}
\scalebox{0.5}[0.5]{
\includegraphics[angle=0]{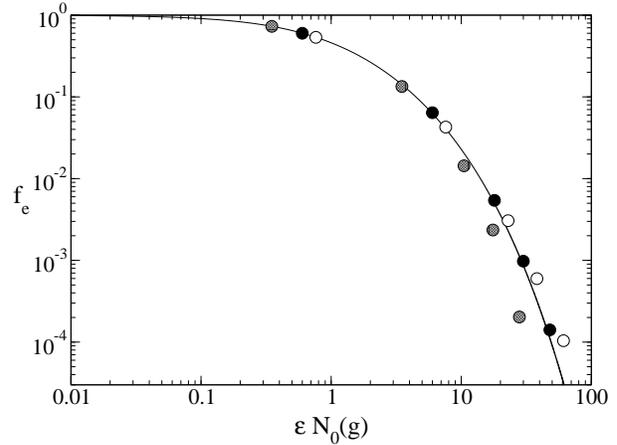} }
\caption{{\bf 1-D Escape Fractions, Arbitrary $g$}  The simulated escape fraction from the middle of a 1-D finite line, for scattering with several values of $g$, over a range of albedos $\epsilon$.  In all simulations, the center to edge extinction (scattering + absorption) optical depth is constant, $\tau=10$.  Three different values of $g$ were simulated (circles);  from darkest to lightest $g=0, \ 1/2, \ -1/3$.  We computed $\N_0(g)$ for each value of $g$:  $\N_0(0)=60.0$, $\N_0(1/2)=35.0$, and $\N_0(-1/3)=76.3$.  For each value of $g$,  there are five different values of $\epsilon$;  from left to right, $\epsilon=0.01, \ 0.1, \ 0.3, \ 0.5, \ 0.8$.  The line is given by eqn. (\ref{fesc_g=0_N0}).}
\label{figure_1Dfesc}
\end{center}
\vspace{0cm}
\end{figure}

The escape fraction for arbitrary $g$ can be approximated by the $g=0$ formula, eqn. (\ref{fesc_g=0_N0}), as we demonstrate next.  Define $n^*$ to be the average number of interactions required for a 50\% chance of back-scattering.  For interactions with scattering parameter $g$, the probability of a forward scatter is $(1+g)/2$ and that of a back-scatter is $(1-g)/2$.  Therefore $n^*$ is defined to satisfy
\begin{equation}
\frac{1}{2}\equiv \sum_{n=0}^{n^*-1} \left(\frac{1-g}{2}\right)\left(\frac{1+g}{2}\right)^n\ ,
\end{equation}
which leads to 
\begin{equation}
\frac{1}{2}=\left(\frac{1+g}{2}\right)^{n^*}\ ,
\end{equation}
or equivalently
\begin{equation}\label{nstar_g}
n^*=\left(1-\frac{\ln(1+g)}{\ln 2}\right)^{-1}\ .
\end{equation}
Every $n^*$ interactions acts like a single $g=0$ interaction.  The probability of absorption after $n^*$ interactions is $1-(1-\epsilon)^{n^*}$.   Consequently, the escape fraction is approximately the same as the $g=0$ formula, eqn. (\ref{fesc_g=0_N0}), with a re-scaled $\N_0$ and absorption albedo, $\N_0^*$ and $\epsilon^*$, given by
\begin{eqnarray}\label{rescalings}
\N_0^*&\equiv&\N_0/n^* \\
\epsilon^*&\equiv&1-(1-\epsilon)^{n^*}\ .
\end{eqnarray}
The approximate escape fraction for arbitrary $g$ is 
\begin{equation}
\fesc=1/\cosh\left(\sqrt{Y}\right)
\end{equation}
with
\begin{eqnarray}
Y&=&2\N_0^*\epsilon^* \nonumber \\
&=&2\frac{\N_0}{n^*}\left(1-(1-\epsilon)^{n^*}\right)\ ,
\end{eqnarray}
where $\N_0=\N_0(g)$ is calculated for the given value of $g$ and $n^*$ is a function of $g$ through eqn. (\ref{nstar_g}).  Note that for 1-D transfer when $g=0$, then $\N_0(g=0)=\frac{1}{2}(\tau^2+2\tau)$, which we have verified with simulations.  If $n^*<\N_0$ then there is enough scatterings for this approximation to hold.  If  $\epsilon n^*\ll 1$ also holds, then $\epsilon^*\approx n^*\epsilon$ and the rescaling leaves the product $\epsilon\N_0$ unchanged.  In this limit, the escape fraction given in eqn. (\ref{fesc_g=0_N0}) is valid for any type of scattering, and represents, therefore, the generic escape fraction form for any type of ``random walk'' photon transfer.  On the other hand, if $n^*\geq \N_0$ then this approximation breaks down, and the trajectory of the photon is more accurately characterized by straight line motion, with negligible back-scattering, eqn. (\ref{fesc_g=1_N0}).  As shown in Figure \ref{figure_1Dfesc}, the approximation that the escape fraction is given by eqn. (\ref{fesc_g=0_N0}) works well when $\fesc\geq 1\%$, and gives a decent order of magnitude estimate when $\fesc$ is lower (such cases are observationally unimportant).
     
\end{document}